\documentclass[aps,prf,showpacs,floats,onecolumn,floats,superscriptaddress,floatfix,longbibliography]{revtex4}
\usepackage{bm}
\usepackage{hyperref}
\usepackage[usenames,dvipsnames]{xcolor}
\usepackage[normalem]{ulem}
\usepackage[utf8]{inputenc}
\usepackage{amsmath}
\usepackage{amsthm}
\usepackage{amsmath}
\usepackage{amsfonts}
\usepackage{blindtext}
\usepackage[utf8]{inputenc}
\usepackage[T1]{fontenc}
\usepackage{graphicx}
\usepackage{float}
\usepackage{wrapfig}
\usepackage{bm}
\usepackage{braket}
\usepackage{grffile}
\usepackage[normalem]{ulem}
\usepackage[usenames,dvipsnames]{xcolor}
\def\PC#1{{\textcolor{black}{ #1}}}    
    
\def\RMV#1{{}}

\newcommand{\beq}{\begin{equation}}
\newcommand{\eeq}{\end{equation}}

\newcommand{\bk}{\bm{k}}
\newcommand{\cI}{{\cal I}}

\newcommand{\bx}{\bm{x}}
\newcommand{\by}{\bm{y}}
\newcommand{\br}{\bm{r}}

\newcommand{\bv}{\bm{v}}

\newcommand{\Lcal}{\mathcal{L}}

\begin{document}

\title{Reconstruction of turbulent data with deep generative models for semantic inpainting from TURB-Rot database} 
\affiliation{Dept. Physics and INFN, University of Rome ``Tor Vergata'', Italy.}
\affiliation{Center for Life Nano Science@La Sapienza, Istituto Italiano di Tecnologia and INFN, University of Rome ``Tor Vergata''}
\affiliation{Department of Mechanical Engineering, Johns Hopkins
University, Baltimore, Maryland 21218, USA.}
\author{M. Buzzicotti$^1$, F. Bonaccorso$^{1,2}$,  P. Clark Di Leoni$^3$, L. Biferale$^1$}

\date{\today}
\begin{abstract} 
We study the applicability of tools developed by the computer vision community for {\it feature learning} and {\it semantic image inpainting}  to perform  data reconstruction  of fluid turbulence configurations. The aim is twofold. First, we explore on a quantitative basis, the capability  of Convolutional Neural Networks  embedded in a Deep Generative Adversarial Model (Deep-GAN) to generate missing data in turbulence, a paradigmatic high dimensional chaotic system. In particular, we investigate their use in reconstructing two-dimensional {\it damaged} snapshots extracted from a large database of numerical configurations of 3d turbulence in the presence of rotation, a case with multi-scale random features where  both large-scale organised structures and small-scale highly intermittent and non-Gaussian fluctuations are present. Second, following a reverse engineering approach, we aim to rank the input flow properties (features) in terms of their qualitative and quantitative importance to obtain a better set of reconstructed fields. We present two approaches both based on Context Encoders. The first one infers the missing data via a minimization of the $L_2$ pixel-wise reconstruction loss, plus a small adversarial penalisation. The second, searches for the closest encoding of the corrupted flow configuration from a previously trained generator. Finally, we  present a comparison with a different data assimilation tool, based on Nudging,  an {\it equation-informed} unbiased protocol, well known in the numerical weather prediction community. The  TURB-Rot database, \url{http://smart-turb.roma2.infn.it}, of roughly 300K 2d turbulent images is released and details on how to download it are given.    
\end {abstract}

\maketitle

\section{Introduction}
Data assimilation (DA) is the art of interpolating, de-noising, super-resolving or filling missing information from a single realization, a time or oriented series, or a statistical sample of some random or deterministic dataset \cite{little2019statistical,  asch_data_2016}. It is important in many areas, such as classification and inpainting in computer vision (CV) \cite{pathak2016context,yeh2017semantic,ulyanov2018deep}, developing natural language processing \cite{mikolov2013distributed,bowman2015large}, inverse problems such as source localization in fluid dynamics \cite{wang_spatial_2019,mons_kriging-enhanced_2019}, or to prepare more and more refined initial conditions in numerical weather predictions \cite{Kalnay,reichstein2019deep,Carrassi18,Bauer15}, just to cite a few examples from different fields. The fact is that our visual and/or data samples are  very diverse, but also highly structured. Sometimes, the naked eye is better than sophisticated algorithms in interpolating, classifying, and guessing the meaning and contents of blurry and  gappy images, if trained enough. It can be a picture taken from a dataset of digitalized images \cite{deng2009imagenet,deng2012mnist,krizhevsky2012imagenet,liu2015deep,netzer2011reading,krause20133d} or a visualization of real physical fields contents, e.g. the flow configuration at a given instant of our atmosphere. The problems are  still the same, we often need to quickly understand the context, to control the quantitative details and contents, to classify it,  to react or to predict the future evolution.
\begin{figure*}%
    \includegraphics[width=1.0\textwidth]{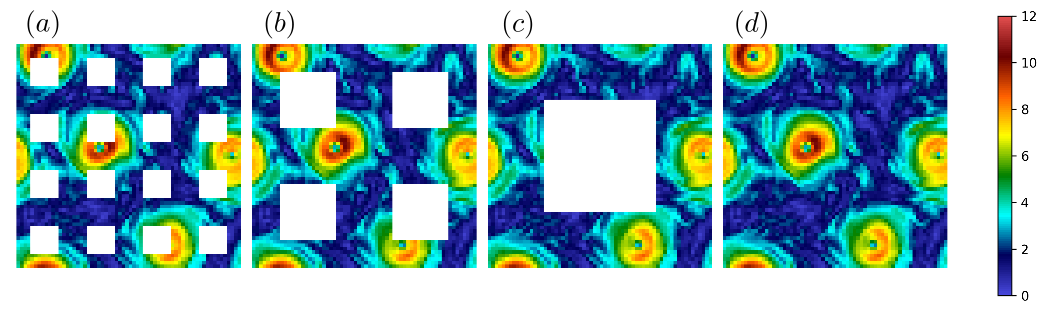}   
    \caption{Example of a typical inpainting exercise on a  2d slice of a 3d turbulent rotating flows (here the color code is proportional to the local velocity intensity). We imagine to have either many small-scale  damages (panel a) or larger and larger gaps (panels b and c) leading to the  need to guess  global semantic information of the image, e.g. the internal structure of a whole vortex and the shear flow in the intermediate regions. The final task is to reconstruct the ground truth (panel d). }
    \label{fig:fig1}
\end{figure*}
For physical applications the challenges, as well as their importance, are clear. The problem can roughly be split into two sub-domains. The case where the missing or blurred information is scattered everywhere but locally affecting only a small cluster of pixels, or where big chunks of data are affected and one cannot use  local interpolation techniques. Consider the three damaged images  of a velocity amplitude on a  2d slice taken from a 3d direct numerical simulations of a rotating turbulent flow shown in Fig. \ref{fig:fig1}. Filling case (a) when the {\it noise} is affecting only small-scale properties might  be relatively simple (and indeed it is, if the gap is smaller than, or comparable with, the smallest active chaotic scale in the flow \cite{meneveau} ). For cases (b) and (c) where the gap is bigger and bigger, it is not at all obvious to guess the ground truth (panel d). In the presence of big holes, we need first to learn the {\it  context}, i.e. the multi-scale coherent structures and their statistical significance, and then inferring what the missing information should be, on the basis of our {\it semantic} understanding and with little possibility to get clues from nearby data. 

In this paper, we aim at investigating these kind of questions using TURB-Rot, a huge dataset  made of hundred thousands of 2d configurations extracted from a direct numerical simulation of 3d turbulent flow in the presence of rotation, see appendix~\ref{appdx:DataBase} and \cite{turbrot}. {Rotating turbulence is a paradigmatic case with very rich physics, where energy injected from the external forcing produces both large-scale cyclonic and anti-cyclonic structures as well as small-scale intermittent homogeneous and isotropic highly non-Gaussian fluctuations, resulting on chaotic flow excitation spanning many decades in scales and frequencies (see, e.g. the Fourier spectrum in Fig.~\ref{fig:rotspectrum} and  \cite{alexakis2018cascades} for a recent review). In our application,  we work in a range of scales where the image is rough and non-differentiable with Holder continuity close to $1/3$ \cite{frisch1995turbulence},  making the problem  much different from the one of a non-linear {\it local} fit.}  Rotating (and stratified) turbulence is also a key set-up in many geophysical applications, both for atmosphere and ocean dynamics. {The flow used and the type of reconstruction problem posed (see Fig.~\ref{fig:fig1}) are interesting {\it per se} and  inspired by the difficulties that arise when trying to assimilate satellite data into weather models where cloud coverage may mask parts of the observed field \cite{Mcnally03} and is partly responsible for up to $75\%$ of satellite data being discarded \cite{Bauer10}.} To achieve the task of reconstructing the missing information  we will make use of two different Machine Learning (ML) algorithms both based on {\it Context Encoders} (CE). The first one was proposed in  \cite{pathak2016context} and consists of two networks. A Generator (G) made of an encoder that stores the context of the input data in a compact (low-dimensional) latent representation by doing a down-sampling of the original corrupted image, and a decoder that builds  on this representation to generate the missing information. The second one is given by a variation of it \cite{yeh2017semantic}. Here, we  first  train a Deep-GAN conditioned on a noise vector, able to produce an entire realistic sample for the flow fields in the whole 2d domain. Then we freeze the whole structure and perform a second optimization with {\it back propagation} on the input noise vector, meant to find the optimal entry data  which generate the known context.  We present results qlso using an equation-informed DA algorithm based on Nudging \cite{Hoke76, Lakshmivarahan13, clark_di_leoni_synchronization_2020}, a method based the Navier-Stokes equations (NSE) with a Newton relaxation feedback term,  imposing a linear reward/penalisation depending if the evolved fields are close/far from the data to be assimilated.

\begin{figure*}
\includegraphics[width=0.8\textwidth]{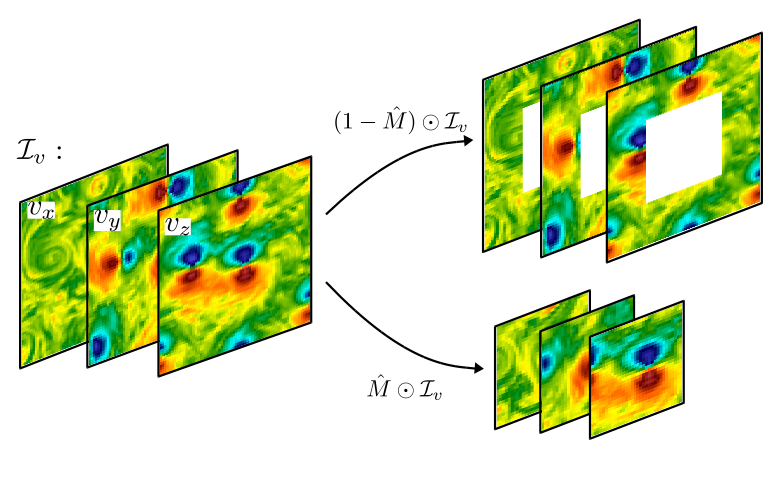}   
    \caption{Sketch of the typical image damaging protocol we implement, separating the set of the velocity components $\cI_v = (v_1,v_2,v_3)$ in the 2d plane $(x_1,x_2)$ in a corrupted image, $(1-\hat{M}) \odot \cI_v $, and in its removed hole,  $\hat{M} \odot \cI_v $, where $\hat{M}$ is a mask made of $0$s (in the external frame)  and $1$s (in the central hole). }
    \label{fig:fig3new}
\end{figure*}

Before moving to the results, let us make a few statements. The literature using equation-informed or equation-free (with or without some physics constraint) tools to model, reconstruct and assimilate fluid flow is huge, one can find some past and recent reviews and books here \cite{little2019statistical,berkooz1993proper,brunton2019data,duriez2017machine,brunton2020machine,Kalnay,Carrassi18,asch_data_2016}. It is also important to point out a few recent developments in the field of turbulence reconstruction  \cite{fukami_super-resolution_2019, fukami_synthetic_2019, callaham_robust_2019, raissi_hidden_2020, clark_di_leoni_synchronization_2020} and in the efforts to merge ML and DA \cite{Brajard19,wikner_combining_2020}. The goal of this paper is not to develop new ML algorithms for DA  neither to spend millions of computing hours to find the optimal combination of network hyper-parameters to improve the final reconstruction. These are important issues that would deserve a different work. {Here we are motivated by exploring how much some of the most popular and recent CV algorithms can be used for a quantitative DA of fluid configurations. We do this by  implementing a series of systematic quantitative benchmarks based on  (i)  spectral properties; (ii) point-wise $L_2$ errors  and (iii) probability distribution functions for large-scale quantities (velocity components) and small-scale ones (vorticity), (iv) sub-grid energy transfer and (v) velocity structure functions.} Similarly, our goal is not meant to explore other subtle and important points connected to the need of imposing soft or hard physical constraints coming from the original properties of the chaotic ODEs or PDEs  \cite{raissi2019physics,wu2020enforcing,wang2019towards,erichson2019physics,lusch2018deep,kashinath2020enforcing,mohan2020embedding}. We limit in the first part to a CV approach, as if we do not know the equations of motion (as it happens in many real applications also in basic science).
Here, we specified  to 2d snapshots from rotating turbulent flows but there is nothing that prevents us (except eventually an issue connected with the capacity of the Deep-GAN structures) to apply the same techniques to other flow configurations and/or fully 3d or to (3+1)d data structures  \cite{mohan2019compressed,wiewel2019latent,kim2019deep,yang_highly-scalable_2019}. Work in this direction is underway and it will be reported elsewhere.
In a final part of the paper, we also address one example from the other -opposite- point of attack: supposing we know and we use the equations of motion and/or the embedded time-evolution. {There are many DA approaches which have been developed in the past for fluid flows, e.g. Kalman filters \cite{Evensen,Houtekamer16}, Kriging \cite{stein2012interpolation,gunes2006gappy}, proper orthogonal decomposition \cite{berkooz1993proper,venturi2004gappy,brunton2019data,scherlrobust} to cite just a few. At difference from what we do here, these techniques are generally applied to repair datasets with a low percentage of missing data, with high temporal sampling frequency or when turbulence is not fully developed \cite{gunes2006gappy,murray2007application}. 
It is also interesting to cite that there exists images and video inpainting approaches based on PDE \cite{schonlieb2015partial}, which search for stationary solutions of 2D NSE to propagate isophote lines from the boundaries into the missing regions \cite{bertalmio2001navier,ebrahimi2013navier}, no applications to space-time fluid configuration have ever been presented up to now.
} As already mentioned above, to compare with ML we chose an algorithm based on Nudging. Finally, we also made available the whole database used for the training and validation of the two  CE1 and CE2, hoping to trigger the interest of the community to improve on our results. \\
The main results of our work can be summarized as follows. 
\begin{itemize}
     \item Both CE1 and CE2 are capable to refill well the missing regions in the corrupted flow configurations, with average local $L_2$ errors that can be as small as $15\%$ for CE1 in the presence of large gaps and $3-4 \%$ for smaller damaged spots. For CE2 we get roughly the double. \item  Multi-scale spectral properties and probability  distribution functions for both velocity and vorticity are extremely well reproduced, including extreme events and sub-grid energy transfer. 
     \item  Concerning features ranking, CE works better with large-scale velocity in  inputs  than with small-scale vorticity  field, probably due to the large intermittent and non-Gaussian properties of the latter. 
     \item  Nudging gives comparable results and the trade-off between the two approaches must be considered depending on the applications and on the quality of the information and numerical tools available, e.g. if we know the equations and we possess the numerical set-up to evolve in time a fully developed 3d turbulent flow, something that typically requires the evolution of billions of degrees of freedom \cite{frisch1995turbulence}).
\end{itemize}
\noindent
The paper is organized as follows. In Sec. \ref{sec:rotation} we give a brief introduction of the physics of turbulence under rotation in Sec. \ref{sec:CE1} we describe the detail of the CE1 network and we show the results of this approach, in Sec. \ref{sec:CE2} we describe the CE2 network and we discuss the results, in Sec. \ref{sec:Nudging} we introduce and discuss the details of Nudging technique. In Sec. \ref{sec:Conclusion} we conclude and summarize all main results. In appendix~\ref{appdx:ACE1}-\ref{appdx:ACE2} we show the detailed structure of the two networks used. In appendix~\ref{appdx:DataBase} we discuss the original data-base used for the CE1 and CE2 and the protocol to download all data.
 
\section{Rotating Turbulence}
\label{sec:rotation}
\begin{figure}
    \centering
    \includegraphics[width=0.45\textwidth]{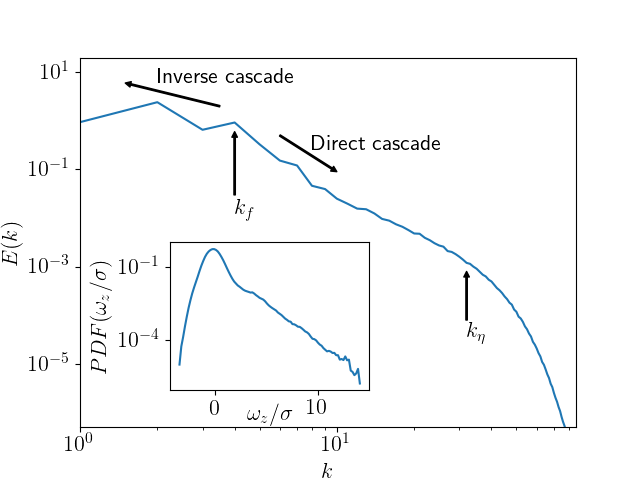}
    \includegraphics[width=0.35\textwidth]{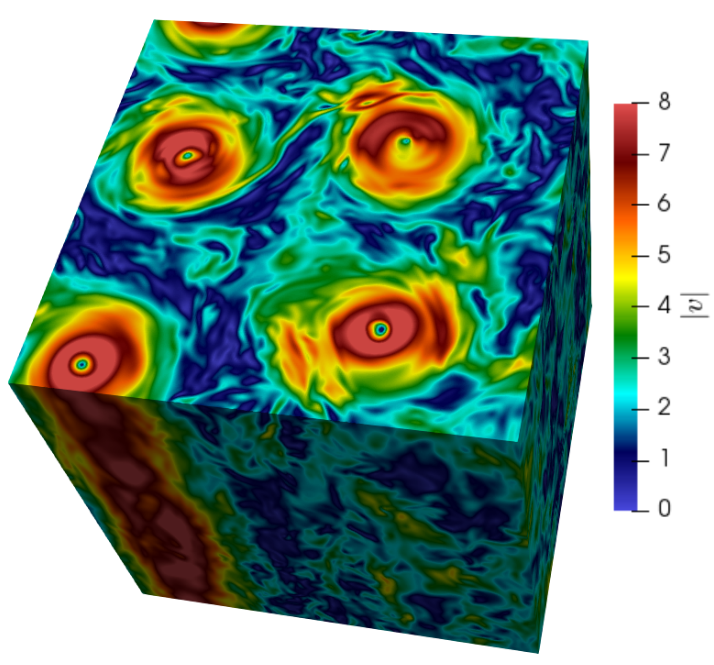}
    \caption{ Right: 3d rendering of the velocity amplitude $|\bv|$, in a rotating turbulent flow. Left:  3d averaged energy spectrum, $E(k) = \sum_{{\bf k}=k}^{k+1} \langle |\hat{\bm{v}}({\bf k})|^2\rangle$, where with $\hat{\bm{v}}({\bf k})$ we denote the Fourier coefficient at wavenumber $\bk$. {The forcing and the Kolmogorov wavenumbers, $k_f, k_\eta$ are indicated with arrows}, notice the presence of power-law energy fluctuations on almost two decades and for wavenumbers smaller and larger than $k_f$, the indication of the split-energy cascade scenario.  Inset: standardized probability distribution function of the vorticity in the direction of the rotation axis, $\omega_3$. Notice the presence of strong non-Gaussian, intermittent and skewed fluctuations up to $O(10)$ (and higher) standard deviations, $\sigma$.}
    \label{fig:rotspectrum}
\end{figure}

Let us first make a small digression introducing the most important features of turbulence under rotation. In a rotating frame of reference, the Navier-Stokes equations take the form:
\begin{equation}
    \frac{ \partial \bm{v}}{\partial t} + \bm{v} \cdot \bm{\nabla} \bm{v} + 2 \Omega \hat{x}_3 \times \bm{v} = - \bm{\nabla} p + \nu \nabla^2 \bm{v} + \bm{f}
    \label{nse}
\end{equation}
where $\nu$ is the kinematic viscosity, $\bm{f}$ is an external forcing mechanism, $2\Omega \hat{x}_3 \times \bm{v}$ is the Coriolis force with $\Omega$ being the rotation frequency and with the rotation axis here chosen to be parallel to $\hat{x}_3$. The incompressibilty condition $\bm{\nabla}\cdot\bm{v}=0$ closes the equations. The most important feature of rotating flows is the tendency to accumulate energy to larger and larger scale, leading to the formation of columnar quasi-2d vortical structures orientated along the rotation axis (see  Fig. \ref{fig:rotspectrum} and \cite{davidson2013turbulence, campagne2014direct, pouquet2013geophysical, yarom2014experimental, sagaut2008homogeneous,buzzicotti2018energy,biferale2016coherent,smith1999transfer}). The phenomenon is accompanied by a simultaneous transfer of energy to smaller and smaller scales too, producing strong non-Gaussian tails in the vorticity probability distribution function (Fig. \ref{fig:rotspectrum}, inset of left panel). We are in the presence of a split-cascade scenario, where the energy injected by the external stirring mechanism is redistributed to both small and large wavenumbers  \cite{alexakis2018cascades}.

The data used in the CV experiments comes from solving Eq.~\eqref{nse} on triple periodic cubic box of size $2\pi\times2\pi\times2\pi$ with $256$ grid points in each direction. {As customary and for the sake of simplicity we used a Gaussian forcing process, delta-correlated in time, allowing to control exactly the energy input and with a support in wavenumber space around $k_f=4$.} We fixed $\Omega=8$, resulting in a Rossby number $Ro = E_{tot}^{1/2}/k_f \Omega\sim 0.1$, where $E_{tot}$ is the flow kinetic energy. The dissipation is modeled by an hyperviscous term $\nu \nabla^{4} \bm{u}$, which replaces the laplacian in \eqref{nse}, with  $\nu = 1.6\times10^{-6}$. A large scale friction term, $-\beta \bm{v}$ is also added to the right hand side of the equations in order to reach stationarity, with $\beta=0.1$. In Fig.~\ref{fig:rotspectrum} (left) we show the energy spectrum $E(k)$ of the simulation, with an arrow denoting the scale where forcing is acting. It is easy to see how both ends of the spectrum get populated with active modes on  almost two decades in  wave-numbers. The inset of the figure shows the probability density function (PDF) of $\omega_3$, the vorticity component parallel to the rotation axis. The distribution is highly skewed and non-Gaussian,  favoring the alignment of vorticity with the rotation axis. Lastly, in the right panel of Fig.~\ref{fig:rotspectrum} we show a visualization the velocity amplitude $|\bv|$ from where we can see the formation of large columnar vortices that cover the whole domain. 

As shown in Fig.~\ref{fig:fig1}, for our inpainting experiments we do not use the whole three dimensional fields, but only two dimensional horizontal cuts in the plane $(x_1,x_2)$. Furthermore, in order to decrease the computational cost of training the two CEs, the velocity images are downsized to $L \times L$ with $L=64$, {by applying a pass-band filter for $k < k_\eta$, where with  $k_\eta \sim 32$ we indicate the Kolmogorov dissipative wavenumber, defined as the scale such that for  $k >k_\eta$ the velocity field becomes smooth \cite{frisch1995turbulence}}. The above phenomenology show the whole complexity of our task: we want to reconstruct/assimilate data from 2d snapshots of 3d highly multi-scale and chaotic fields, characterized by strong intermittent and non-differentiable small-scale fluctuations superposed to  big  large-scale cyclonic and anti-cyclonic vortical structures. These are complex features strongly different from the ones characterising other inpainting tasks based on popular images data-base \cite{deng2009imagenet,deng2012mnist,krizhevsky2012imagenet,liu2015deep,netzer2011reading,krause20133d}.


\section{Context Encoder for Image Generation (CE1)}
\label{sec:CE1}

\begin{figure}%
    \includegraphics[width=0.8\textwidth]{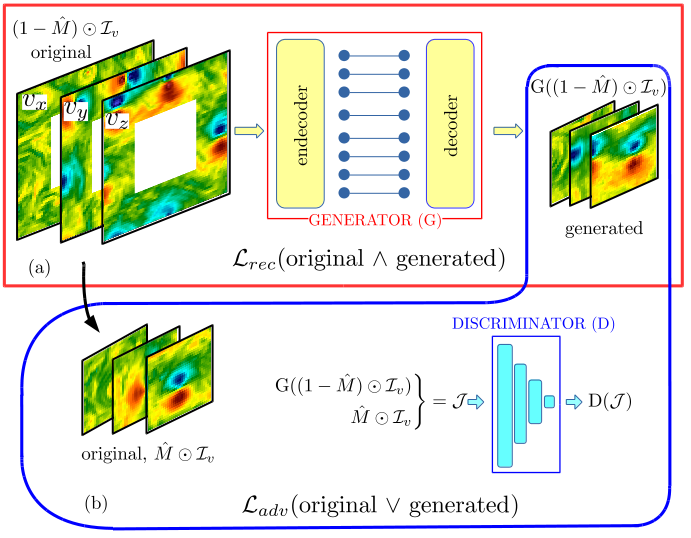}   
     \caption{Sketch of the Context Encoder (CE1) configuration. The machine is made of two blocks. A Generator (G), depicted in panel (a), where we give in input the three channels made of the three corrupted configurations of size $L \times L$ with $L=64$ for the three velocity components in the plane, $v_1(x_1,x_2),v_2(x_1,x_2),v_3(x_1,x_2)$ and obtain as output the fields proposed to fill the gap of size $\ell \times \ell$. The second block is given by a Discriminator (D), see panel (b), which is trained to distinguish the original fields from the ones proposed by the generator for the hole region. See appendices~\ref{appdx:ACE1} and~\ref{appdx:ACE2} for details about the internal structure of G and D.}
    \label{fig:fig2}
\end{figure}

Let us first present the context encoder as suggested in \cite{pathak2016context} with reference to Figs.~\ref{fig:fig3new} and \ref{fig:fig2}  and to appendix~\ref{appdx:ACE1} for specific details. The idea is to have a Deep-GAN that reconstructs the missing part of the three velocity configurations. To that end we define the masking operator $\hat M$, which takes the value $1$ on the missing  pixels  and $0$ on the unaltered locations in our original image $\cI_v$. We will then denote with $\hat{M} \odot \cI_v$ the damaged part of $\cI_v$ and with   $(1-\hat{M})\odot \cI_v$ the region of the flow where we have the full information, see Fig.~\ref{fig:fig3new} for a graphical representation applied to a  set of three velocity components, $\cI_v = (v_1, v_{2}, v_3)$.  The overall architecture is based on a pipeline with first the generator (G) made of  an {\it encoder} which takes as input the damaged image, $(1-\hat{M}) \odot \cI_v$, produces a {\it latent} representation of the relevant features  with a down-sampling of the dataset, then in the second part of the network  a decoder learns how to generalise the features and proposing a candidates for the three velocity components in the missing gap, $G[(1-\hat{M}) \odot \cI_v]$. The Generator part of the context encoder, panel (a) of Fig. \ref{fig:fig2}, is trained  to minimize the  $L_2$ norm between the generated image and the ground truth inside the hole: 
\begin{equation}
\label{eq:Lrec}
\Lcal_{rec} =  \mathbb{E}_{{\cI_v}} \{|| \hat{M}\odot \cI_v  - G[(1-\hat{M}) \odot \cI_v]  ||_2\},
\end{equation}
  where  $G[\bullet]$ indicates the snapshots of three velocity components provided by the network as output and with $ \mathbb{E}_{{\cI_v}} $ we intend average over the set of training configurations. The  final network  is supplemented with an adversarial structure, see panel (b) of Fig. \ref{fig:fig2},  made of a discriminator (D) to provide loss gradients to the context generative encoder G.  The learning protocol is a two-player game where the discriminator D is inputted alternatively with ground truth  images in the hole or the ones generated by the context encoder G and tries to distinguish among them while the generator G tries to confuse D by proposing  samples that appear as close as possible to real turbulent configuration in the missing region.  The adversarial task for the discriminator component of the  encoder is to {\it maximize} the binary cross-entropy function: 
  \begin{equation}
  \Lcal_{adv} =  \mathbb{E}_{{\cI_v}} \{ \log(D[\hat{M} \odot \cI_v]) + \log(1-D[G[(1-\hat{M}) \odot \cI_v])\}
\label{eq:discriminator}
    \end{equation}
    where $D[\bullet]$ is the output of the discriminator with values $\in (0,1)$.  The above expression is maximized when the discriminator is able to assign $1$ to the real images and $0$ to the ones proposed by $G$.  We notice that in the second term of the above expression we only use  data proposed by the generator G in the gap of size $\ell\times \ell$ and do not supply the whole image of size  $L\times L$ to not help  the discriminator to exploit  potential discontinuities when passing from the hole filling to the external frame. \\ Finally, the total loss that must be {\it minimized} to train the Generator part of the  CE1 is given by a linear superposition of the two components: 
      \begin{equation}
 \Lcal = \lambda_{rec} \Lcal_{rec} + \lambda_{adv} \Lcal_{adv},
\label{eq:original}
    \end{equation}
    with $\lambda_{rec}+\lambda_{adv}=1$, and where the adversarial loss is here evaluated only with the images produced by the Generator: $ \log(1-D[G[(1-\hat{M}) \odot \cI_v]).$
    In the following section  we will present the results by varying the typical length scale, $\ell$ of the spatial hole where data are missing  but keeping the overall damaged area the same  (see section \ref{sec:fr1}) or  by varying the  structure of the input data, using a multi-channel information made of three images of the  three velocity components, $v_1,v_2,v_3$ or  a single channel given by  the vertical vorticity $\omega_3$ (see  section \ref{sec:fr2}).

\subsection{Image inpainting: large vs small scales information}
\label{sec:fr1}
\begin{figure}%
    \includegraphics[width=0.8\textwidth]{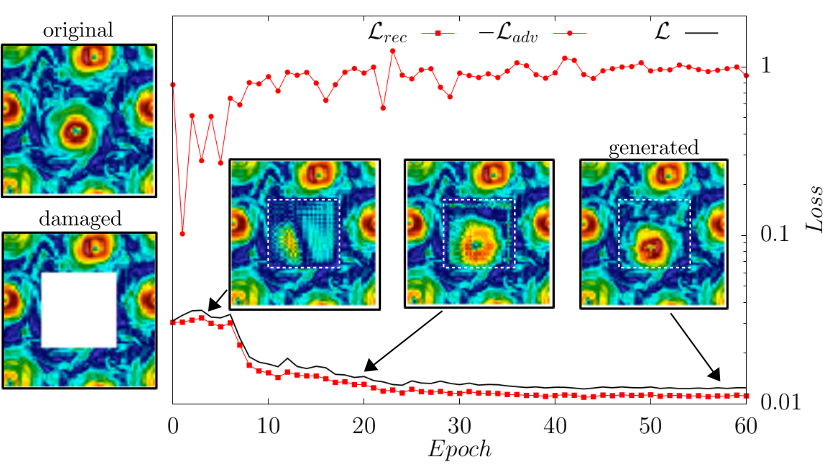}   
    \caption{Example of a training section for CE1. We show the evolution of $\Lcal$, $\Lcal_{rec}$ and $-\Lcal_{adv}$, measured on the validation dataset, as a function of the epoch as well as the reconstructed configuration for $ |\bv|$ at three different steps during training. Notice the tendency to supply better and better inpainting by advancing the learning protocol. 
    The two pictures on the left represent the ground truth in the whole plane and the  damaged input.}
    \label{fig:CE1training}
\end{figure}

We start by analysing the training performance and the final validation for the CE1 network by removing from the input a squared hole of size $\ell \times \ell$ with $\ell=L/2$  located at the centre of the image. Here we present data when we used in input three channels with the three velocity components in the plane and we later compare with the case when the total missing area is the same but each single missing spot is smaller.\\
In Fig. \ref{fig:CE1training} we show the evolution of the reconstruction loss $\Lcal_{rec}$, of the adversarial loss $\Lcal_{adv}$ and of their weighted sum (\ref{eq:original}) obtained with $\lambda_{adv}=0.001$, measured on the validation dataset, as a function of the epochs during training, for details see appendix~\ref{appdx:ACE1}. 
For training we have used a total number of configurations $N_{tra}=81920$ while for validation we used a set of $N_{val}=20480$ configurations never seen during training. As one can see from Fig. \ref{fig:CE1training}, the training converges rapidly without any over-fitting. {Indeed, to improve the network generalization we have added a dropout of $20\%$ on the `features vector' in CE1, see appendix A. The key idea behind dropout, is to randomly remove units (along with their connections) from the neural network during training. This prevents units from adapting too much to the training dataset and it reduces significantly over-fitting \cite{srivastava2014dropout}.}
In the inset of the same Fig. \ref{fig:CE1training} we show the quality of the gap filling  at different epochs, to give a visual demonstration of the ability of CE1 to progressively learn the correct flow structures. {Each mini-batch used in our training is composed of $128$ images and each epoch consists of $640$ mini-batches (see appendix A for more details). Training has been performed also with a smaller number of mini-batches but obtaining worse results. Hyper-fine optimization in the size of each mini-batch has not been performed because not our main interest here.}  \\

\begin{figure}%
    \includegraphics[width=1.0\textwidth]{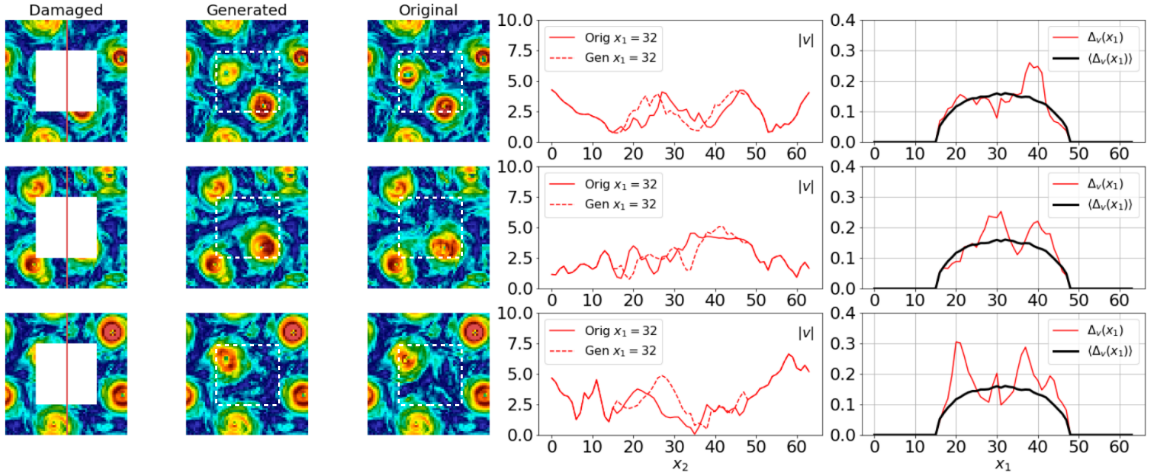}   
    \caption{ Validation of gap filling by CE1 for three different generated fields (one for each row). Images show the magnitude of local velocity field, $|\bv(\bx)|$. 1st column: damaged image in input. 2nd column: image generated in output. 3rd column: ground truth. 4th column: generated (dashed) and ground truth (solid) profiles of $|\bv(\bx)|$ along the  vertical line shown in the 1st column. 5th column: $L_2$ error, $\Delta_v(x_1)$ vs $x_1$, given by expression (\ref{eq:error})  for each image (red line) and the  average error $\langle \Delta_v(x_1) \rangle_N$  over 
    $N=2048$ different 
     images (black curve). }
    \label{fig:CE1spotsvalidationb}
\end{figure}

We now proceed with a more quantitative analysis of the CE1 performances using the network frozen at its final epoch configuration.  In Fig.~\ref{fig:CE1spotsvalidationb} we show a series of 3 different gap filling experiments using configuration never showed during training as well as a point-wise analysis of the normalized  $L_2$ norm calculated on vertical lines for each fixed horizontal position, $x_1$:
\beq
\Delta_v(x_1)  =\frac{{\ell^{-1}\sum_{x_2} \sum_{i=1}^3 [v^{truth}_i(x_1,x_2) -G_i(x_1,x_2)]^2} }{E_{tot}}
\label{eq:error}
\eeq
where $G_i(x_1,x_2)$ for $i=1,2,3$ are the three components of the  field generated by CE1 inside the missing region. The above expression is  normalized dividing by the averaged flow energy over a subset of $N=2048$ of the validation dataset that are the configurations used in our reconstruction experiment: $E_{tot} =  {L^{-2}  \sum_{x_1,x_2} \sum_{i=1}^3 \langle [v^{truth}_i(x_1,x_2)]^2}\rangle_N$, where  we used  $\langle \bullet \rangle_N $ for a shorthand notation of the average. 
\begin{figure}%
    \includegraphics[width=1.0\textwidth]{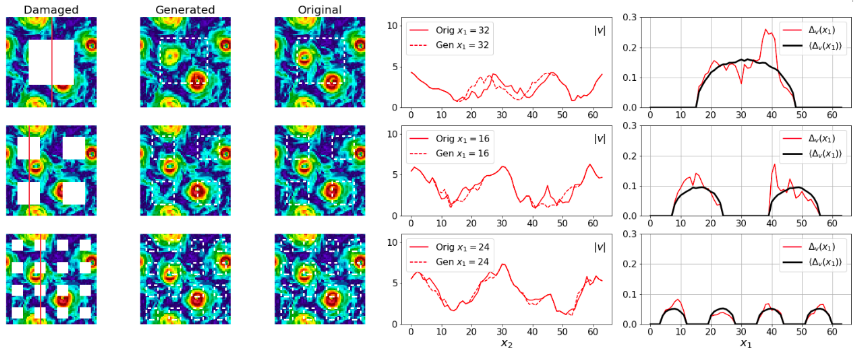}   
    \caption{The same as in Fig.(\ref{fig:CE1spotsvalidationb})  at changing the typical size of the damaged region, $\ell/L=1/2,1/4,1/8$ from top to bottom rows. }
    \label{fig:CE1spotsvalidation}
\end{figure}
It is interesting to notice (5th column, red line) that the local normalized $L_2$ error for each snapshot is of the order of $10-20\%$, when we are in the middle of the missing region, showing a very good reconstruction property of our CE. In the same panels we plot  the averaged errors, $\langle \Delta_v(x_1)\rangle_N$. As on can see, the max averaged error is $O(15) \%$ (black lines). Notice also that the error slightly improve for horizontal position close to the gap edge, indicating that the network is able to detect local correlations also.  
\noindent In Fig.~\ref{fig:CE1spotsvalidation} we test the CE1 at changing the spatial distribution of the missing data, by distributing the same amount of total gaps but on squares with smaller and smaller size $\ell/L=1/2,1/4,1/8$. 
As one can see from the black curves in the last column of the same figure, the averaged
errors, $\langle \Delta_v(x_1)\rangle_N$, become smaller and smaller by decreasing $\ell$, reaching a maximum averaged error as small as $3-4\%$  for $\ell/L=1/8$. {It is important to stress that this is a very good result, being the corrupted size still much bigger than the differentiable limit given by the Kolmogorov scale $\eta = 1/ k_\eta \sim \ell/L =1/64$, (see Fig. \ref{fig:rotspectrum}). In other words, we always work  in the scale range where the fields are  rough and non-differentiable } with Holder continuity close to $1/3$ \cite{frisch1995turbulence},  making the problem  much different from the one of non-linear local fit that other classical inpainting  methods can  handle well \cite{bertalmio2000image,barnes2009patchmatch,efros1999texture,osher2005iterative}. 
In the rightmost panel of Fig.~\ref{fig:physics} we show the most stringent test one can perform, checking the point-by-point reconstruction properties by plotting the  probability 
distribution function of the local $L_2$ error: $$\Delta_v(x_1,x_2)  =\frac{{ \sum_{i=1}^3 [v^{truth}_i(x_1,x_2) -G^{v}_i(x_1,x_2)]^2} }{E_{tot}}.$$ The panel shows that we obtain a pretty good performance to assimilate data even local-wise with very small probability to make big errors. Furthermore, when $\ell/L$ becomes small there is a sharp improvement for the {\it worst case} as shown by the sharp reduction in the right tail extension when comparing  $\ell/L =1/8$ with $1/2$.\\

In Fig.~\ref{fig:physics} we perform also two less stringent statistical test concerning the reconstructed flow. First, in the left panel we show  the  2d spectrum:
\beq
E(k) = \sum_{k < {\bk} <k+1} \langle \hat v_i({\bk}) \hat v_i({-\bk}) \rangle_N 
\label{eq:2dspectrum}
\eeq
where ${\bf k} = (k_1,k_2)$ and we calculated the Fourier transform of the velocity field, $\hat v_i({\bf k})$, over the whole 2d configuration  with the gap filled as proposed by the output of CE1. In the same panel we also show the spectrum calculated on the corrupted input to quantify how much information (energy) we miss at different scales.  Curves here are  averaged over $N=2048$ configurations. In the same figure, middle panel, we show  the probability distribution functions (PDF) for the velocity magnitude, $ v $,  and for the out-of-plane  vorticity, $\omega_3$. All comparisons between generated and the ground truth are very good even for the strongly non-Gaussian distribution  enjoyed by the  $\omega_3$ field. \\
{In Fig.~\ref{fig:physics_sgs}, we focus on the comparison of purely turbulent quantities, namely, the energy transfer across scales $\Pi_{\Delta}$ and the structure functions $S_p(r)$ up to $p=4$. To define the energy transfer in physical space we follow the approach common in large-eddy-simulations (LES) \cite{Pope00, meneveauARFM}, 
\begin{equation}
\label{eq:SGS_Pi_ples}
 \Pi_{\Delta}= \partial_j \bar{v}_i \left ( \overline{v_iv_j} -  \overline{\bar{v}_i\bar{v}_j} \right ) \ ,
\end{equation}
where the $i,j$ indices run over the two velocity components parallel to the plane, namely $(1,2)$. Both coarse-grained velocity components, ${\bar{v}_i}$, are obtained by a convolution operation between the velocity field $v_i(\bx,t)$ and a filter kernel $G_{\Delta}(\bx)$, 
\begin{equation}
{\bar{v}_i}(\bx,t) \equiv \int_S  d\by \ G_\Delta(|\bx-\by|)\ v_i(\by,t),
\end{equation}
where $S$ represents the plane surface. In particular, as filter kernel we have used the `sharp spectral cutoff' in Fourier space \cite{Pope00} with a cutoff $\Delta = \pi/k_c$ and $k_c=16$. 
The main panel of Fig.~\ref{fig:physics_sgs}(a) shows the PDF of $\Pi_{\Delta}$ constrained on the damaged area of the velocity planes, from which we can see the good agreement between original and generated data both showing the same deviation from gaussianity. In the inset of the same figure two visualizations of the energy transfer on the same plane either from the original or the reconstructed fields are presented to appreciate visually the good quality reached in the reconstruction by CE1 network.
In panel (b) of Fig.~\ref{fig:physics_sgs}, in order to estimate the quality of the reconstruction in terms of velocity correlations at different scales, we analyse the statistics of the longitudinal velocity increments defined as $\delta_r v = (\bv(\bx+\br)-\bv(\bx))\cdot \br/r$ \cite{frisch1995turbulence}. In particular in Fig.~\ref{fig:physics_sgs}(b) we compare the 3rd and the 4th order structure functions, defined as;
\begin{equation}
S_p(r) \equiv \langle [\delta_r v]^p\rangle,
\label{eq:SFlong}
\end{equation}
where the $\langle \cdot \rangle$ means an average over the planes and all the $2048$ reconstructed fields. Also from this analysis we found a high agreement between original and reconstructed data. In the main panel we present the log-log plot of $S_4(r)$ data, from which we can see a perfect collapse between the original and generated data. In the inset of the same figure the third order structure function is presented in log-lin scale from which we can appreciate that also the change in the sign of $S_3(r)$ is captured by the CE1 reconstruction.} 

\PC{As a final statistical analysis of the reconstructed images we show, in Fig.~\ref{fig:maxes}, a scatter plot of the extreme values of the velocity (in panel (a)) and vorticity (in panel (b)) coming from the original data and the field generated by CE1 inside the predicted region. There is a very good correlation between the original and reconstructed data for both quantities, indicating that the CE1 is able to properly reproduce the correct maximum value inside each gappy region.}

\begin{figure}%
    \includegraphics[width=1.0\textwidth]{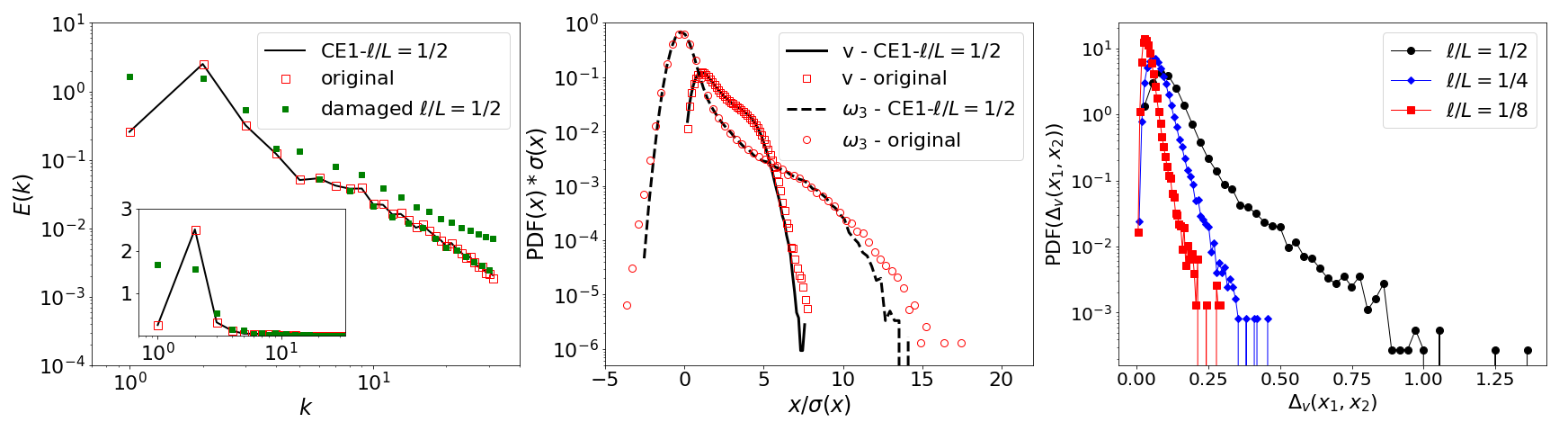}   
    \caption{ Left: comparison between the 2d spectrum (\ref{eq:2dspectrum}) of the ground truth original DNS data in the whole 2d plane and the data reconstructed by the CE1 with $\ell/L=1/2$, as well as the spectrum calculated on the damaged image; the inset shows the same data in log-lin scale. Middle: comparison of standardised PDFs of velocity magnitude, $v=|\bv|$ and vorticity $\omega_3$. Right: Lin-Log  PDF for the point-wise $L_2$ error, $\Delta_v(x_1,x_2)$ at changing $\ell/L$. }
    \label{fig:physics}
\end{figure}
\begin{figure}%
    \includegraphics[width=1.0\textwidth]{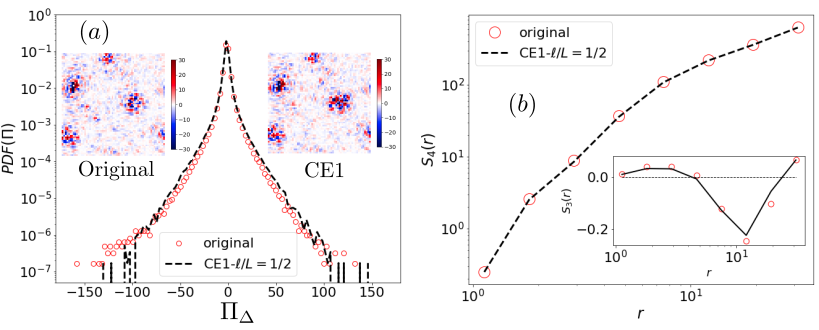}   
    \caption{ {Panel (a): Comparison of the PDF of the energy transfer $\Pi$ measured inside the damaged region from both the original (red circles) and reconstructed data using CE1 network (dashed black line). Panel (b): Comparison of the longitudinal structure function, $S_p(r)$, for both original (red circles) and reconstructed data (dashed black lines). In the main panel $S_4(r)$ is presented in log-log scales, while in the inset the 3rd order structure functions, $S_3(r)$ are presented in log-lin scale.}}
    \label{fig:physics_sgs}
\end{figure}

\begin{figure}%
    \includegraphics[width=0.46\textwidth]{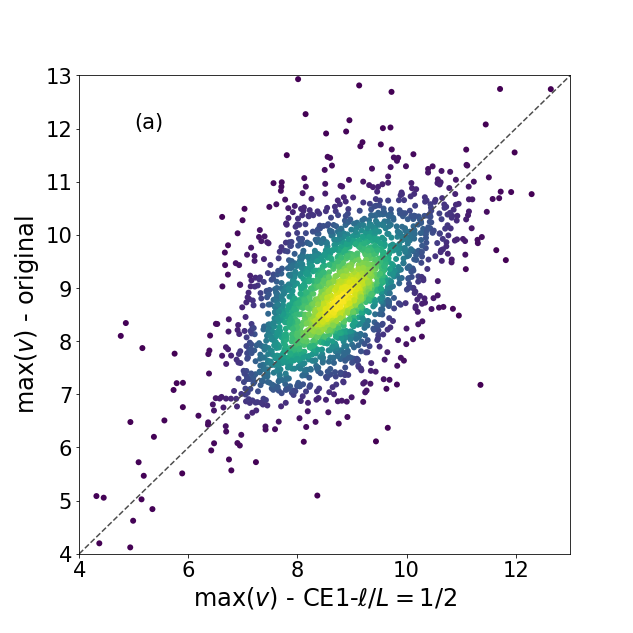}
    \includegraphics[width=0.46\textwidth]{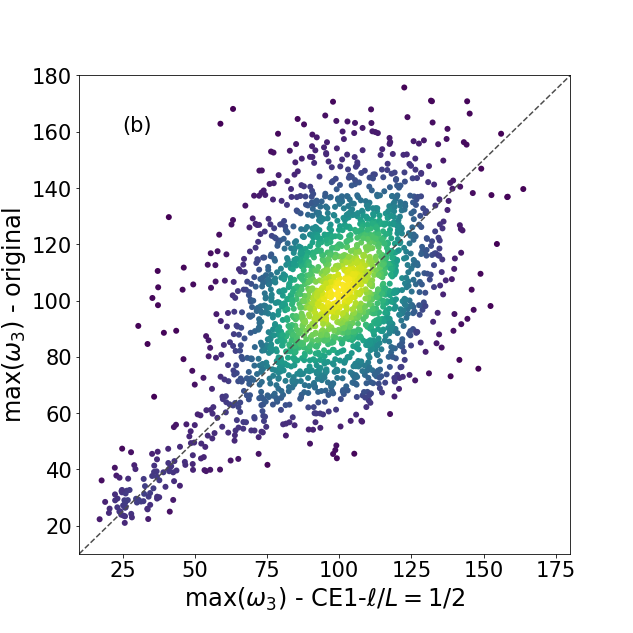}
    \caption{ \PC{Scatter plot of the maximum values inside the reconstructed region of the (a) velocity and (b) vorticity calculated from the original data and the one produced by CE1. Colours are proportional to the density of points in the scatter plot.}}
    \label{fig:maxes}
\end{figure}

\subsection{Features ranking}
\label{sec:fr2}
\begin{figure}%
    \includegraphics[width=1.0\textwidth]{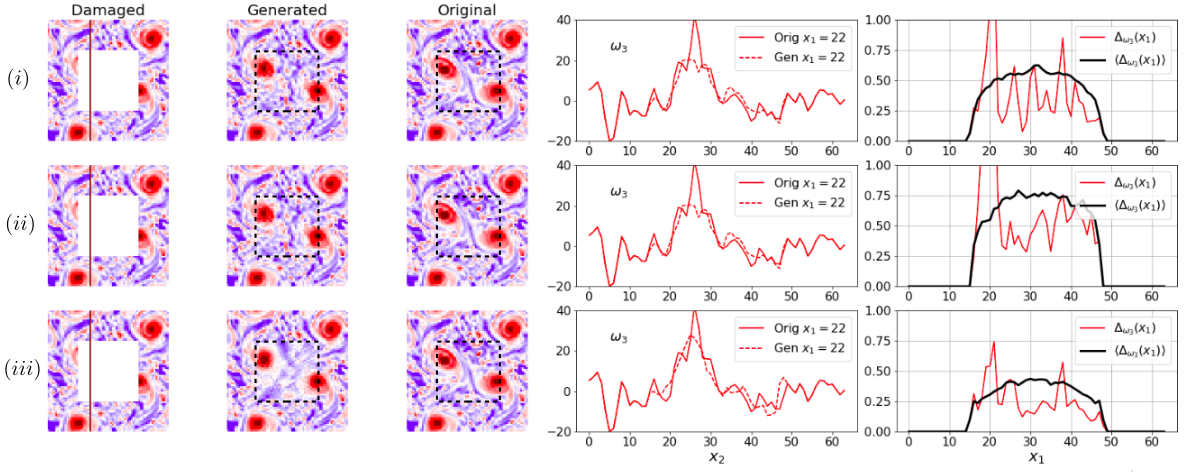}   
    \caption{Vorticity reconstruction with CE1. Comparison for vorticity generation using loss given by case (i), Eq. (\ref{eq:original}) first row; obtained by using vorticity in input and in output as in case (ii), Eq. (\ref{eq:CE1ii}), second row; and for case (iii), Eq. (\ref{eq:CE1iii}) third row. Errors are defined following  Eq. (\ref{eq:error})}
    \label{fig:3channels}
\end{figure}

\begin{figure}%
    \includegraphics[width=1.0\textwidth]{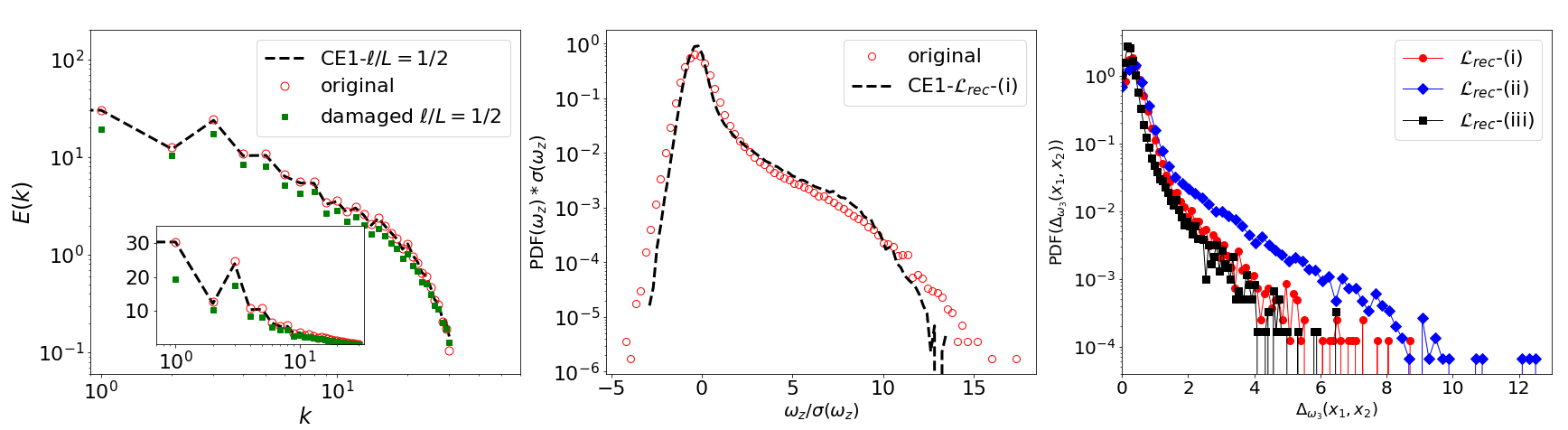}   
    \caption{Left: comparison between spectrum of vorticity for ground truth (original DNS data in the whole 2d plane), for the data reconstructed by the CE1 with $\ell/L=1/2$ using the protocol based on the loss function of case (iii) given by Eq. (\ref{eq:CE1ii}) and for the spectrum calculated on the damaged input images, inset the same but in log-lin scale. Middle: comparison for standardised PDFs of  vorticity $\omega_3$. Right:  Lin-log plot of the PDF for the point-wise $L_2$ error, $\Delta_v(x_1,x_2)$ using the three different CEs trained with the loss functions for cases (i-iii).}
    \label{fig:physicsomega}
\end{figure}

In this section we discuss a different question, connected to use  DA tools as a reverse engineering approach to perform features ranking, i.e. to assess the quality and quantity of the  input data on the basis of the obtained output.  To do that, one can change the set of fields supplied in the CE  input and/or  change the definition of the loss function in order to optimize some features of the generated output. Let us suppose, for example, that we want to reconstruct with high accuracy only the out-of-plane vorticity field, $\omega_3$. One question one can ask is that weather it is better to work with (i) the three velocity channels as input and optimising the reconstruction against the corresponding three velocity outputs in the missing gap, as done in the previous section,  (ii)  $w_3$ as input and output (one channel only) or (iii) using $v_1,v_2,v_3$ as input but adding in the loss also a penalization against bad reconstruction of gradients.   The idea is to test the importance of having both large and small-scales information in the CE1 input/output, i.e. semantic contents that have larger and smaller {\it influence size}. The two above cases labelled with (ii) and (iii) would correspond to the following  loss function: 
\begin{equation}
\Lcal_{rec} = \mathbb{E}_{{\cI_\omega}}  || \hat{M} \odot (\cI_\omega  - G[(1-\hat{M}) \odot \cI_\omega] ) ||_2
\label{eq:CE1ii}
\end{equation}
for case (ii) and to  
\begin{equation}
\Lcal_{rec} =   \alpha  \mathbb{E}_{{\cI_v}} \{|| \hat{M} \odot (\cI_v  - G[(1-\hat{M}) \odot \cI_v] ) ||_2 \} + (1-\alpha) \mathbb{E}_{{\cI_v}} \{|| \hat{M} \odot ( \omega_3 \odot \cI_v  - \omega_3 \odot  G[(1-\hat{M}) \odot \cI_v] ) ||_2\}
\label{eq:CE1iii}
\end{equation}
for case (iii), where with $\omega_3 \odot $ we intend the operation to extract the $\omega_3$ component out of the three $v_1,v_2,v_3$ original fields, $\cI_v$ or of the ones generated by $G[\bullet]$ and we have introduced a parameter $\alpha \in [0:1]$ that weighs the loss due to reconstruction of $\bv$ and  the one due to $\omega_3$. 
In the first row of Fig. (\ref{fig:3channels}) we show the results for  vorticity reconstruction for case (i), with loss given by Eq. (\ref{eq:discriminator}), while in second and third rows we show the same but for cases (ii) and (iii), which correspond to losses given by Eqs. (\ref{eq:CE1ii}) and (\ref{eq:CE1iii}), respectively.    The latter obtained with mixing parameter $\alpha=0.5$. Comparing results from first and second row, $\langle \Delta_{\omega_3}(x_1)\rangle \sim 55\%$ and $75\%$ respectively, we see that in order to have an improved  reconstruction of vorticity it is better to train with the whole set of three velocity field (first row, case i) than using vorticity directly
(second row, case ii). This is somehow natural because for the former we have more information in input than in the latter. Moreover, vorticity is a  highly skewed and non-Gaussian variable \cite{frisch1995turbulence}, properties that could be problematic to handle as input in the network as done in  case (ii). Conversely, by comparing first and third rows, cases (i) and (iii),  we learn that keeping the three velocity in inputs (max information) and penalising the output in terms of the $L_2$ norm for vorticity does help to give a better reconstruction decreasing  $\langle \Delta_{\omega_3}(x_1)\rangle$. Let us notice, however, that the final result is never exceptionally satisfying, with averaged errors in the middle of the missing region that are always around $35\%-40\%$, at the best. From the right panel of Fig.~\ref{fig:physicsomega} we can indeed see that even for the better results, case  (iii), the PDF of the point-wise reconstruction error has  now a pretty extended right tail, indicating the partial failure to have a good local reconstruction, even though the spectral and PDF for the vorticity field are satisfactorily reproduced (left and middle panel).  The difficulty to reconstruct point-wise the correct vorticity properties can be due to different reasons. First, the field is strongly non-Gaussian and with a very small spatial correlation length of the order of $\eta$, hence much smaller than $\ell$. Second, as one can see from the first three columns of Fig. ~\ref{fig:3channels} it is enough a small mismatch for the position of the vorticity peaks between the ground truth and the reconstructed images to produce a big {\it local} error while the overall image is still very close to the original one. Third, vorticity field is obviously much  more sensitive to small details, with a Fourier spectrum  enhanced by a factor $\propto k^2$ wrt the kinetic energy case, something that is particularly critical in the presence of rough non-differentiable fluctuations as in our turbulent case.

\section{Context Encoder with back-propagation to the input data (CE2)}
\label{sec:CE2}
\begin{figure}%
    \includegraphics[width=0.9\textwidth]{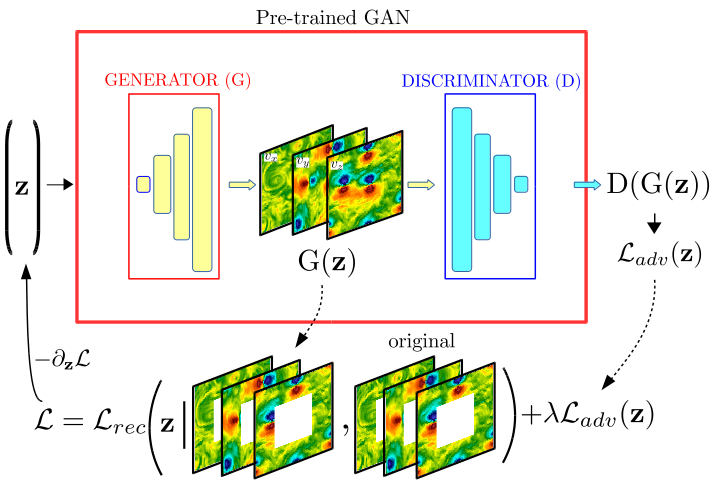}   
    \caption{
    Sketch of the Context Encoder (CE2) configuration. The machine is made of a typical Deep-GAN structure \cite{yeh2017semantic,goodfellow2014generative} which is pre-trained to generate output images on the whole 2d plane, $G[{\bf z}]$ from a random input ${\bf z}$. To apply it for gap-filling, the GAN is frozen and a back-propagation to the input data is implemented. Only ${\bf z}$ is evolved to match the given corrupted image, $(1-\hat{M}) \odot \cI_v$, where we have data.  }
    \label{fig:CE2}
\end{figure}

\begin{figure}%
    \includegraphics[width=1.0\textwidth]{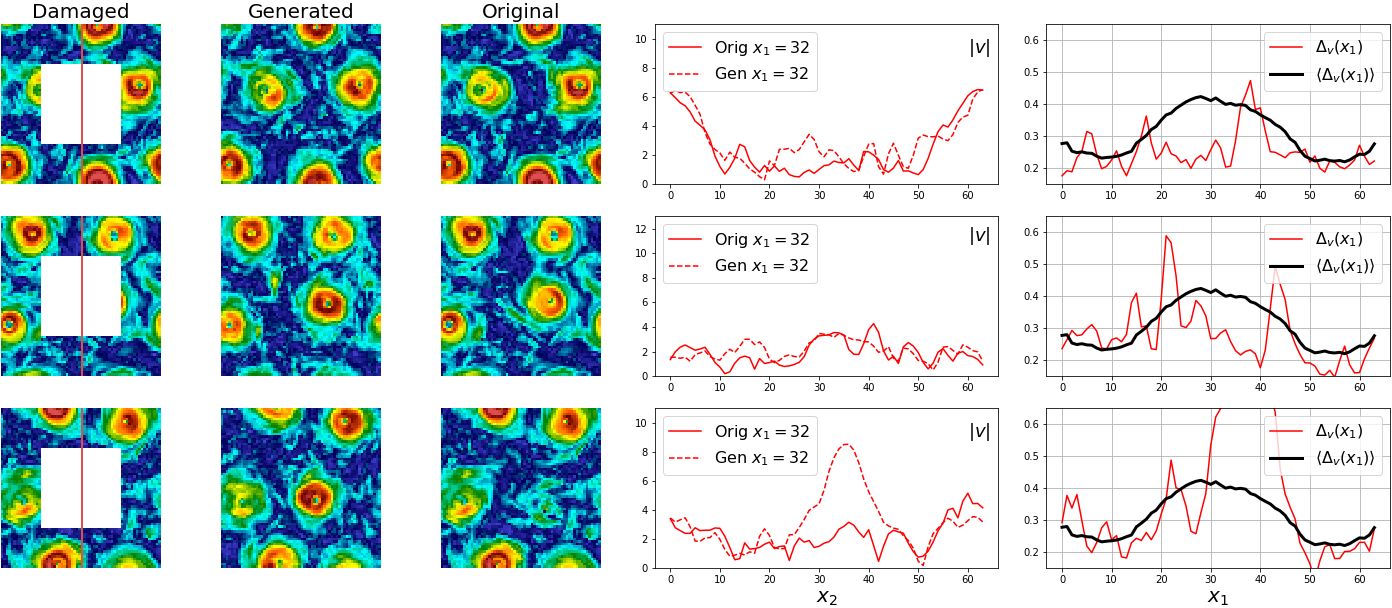}   
    \caption{The same of Fig.~\ref{fig:CE1spotsvalidationb} but for CE2. Notice the 3rd row, where the network puts a vortex in the gap region which is not present in the original case, hence the big error $\Delta_v(x_1)$ for $x_1 \sim 40$.}
    \label{fig:CE2spotsvalidation}
\end{figure}

\begin{figure}%
    \includegraphics[width=1.0\textwidth]{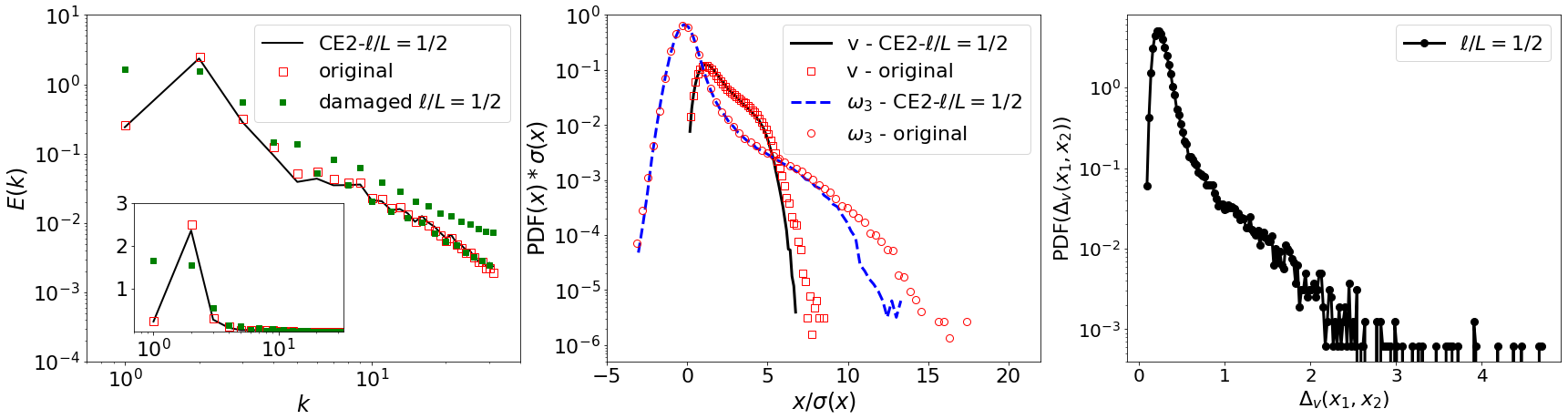}   
    \caption{  The same as in Fig.~\ref{fig:physics} but for CE2. Comparison for spectra (left), PDF of velocity and vorticity (middle).  Right: Lin-log plot of the PDF for the point-wise $L_2$ error calculated in the whole area.}
    \label{fig:physics2}
\end{figure}

The second context encoder we analyze is based on the concept of constrained image generation \cite{yeh2017semantic}. The idea proceeds in two steps. First a GAN is trained to generate uncorrupted flow configurations in the whole 2d domain, see Fig. \ref{fig:CE2}. The generation is conditioned on a set of random input values, ${\bf z} = (z_1,z_2,\dots,z_T)$, with $T=100$ in our case,   which are transformed by the trained machine in realistic flow configurations, at least as realistic as being not distinguishable from the discriminator in the GAN itself.   
Then, the GAN is frozen and for the image reconstruction task a search for the optimal set of ${\bf z}$ inputs is performed, asking  the machine to reproduce with high fidelity only the uncorrupted region. The idea is that once the GAN is trained to generate  the entire turbulent snapshots, we just need to search for the closest configuration that can be generated by the machine and that matches the known data, $(1-\hat{M}) \odot \cI_v$. Moreover, also a penalty against generation of patches that are disconnected by the frame will be added, via the insertion of a discriminator loss (see below and  Fig.~\ref{fig:CE2}). At difference form CE1, here we do not  use the structure of the mask for training. After the Deep-GAN has been trained \cite{goodfellow2014generative,denton2015deep,simonyan2013deep} we proceed to recover the optimal ${\bf z}^*$-encoding
that is closest to the corrupted image. After ${\bf z}^*$ is obtained we can input it to the GAN to obtain the whole configuration also inside the gaps. In practice, in order to fill one damaged field, $(1-\hat{M}) \odot \cI_v$, we obtain ${\bf z}^*$ by {\it minimizing} the combined loss:
\beq
 \Lcal = \Lcal_{rec} + \lambda_{adv} \Lcal_{adv}
 \label{eq:totalCE2}
\eeq
where 
\beq
\Lcal_{rec} = || {(1-\hat M)} \odot (\cI_v  - G[{\bf z}] ) ||_2
\label{eq:lossCE2rec}
\eeq
by back propagating with respect to the input data: $ -\partial_{{\bf z}} \Lcal_{rec} $, where we need to notice that in (\ref{eq:lossCE2rec}) we are using information only on the available data by applying the mask $(1-\hat M)$ to all velocity components. The second term in (\ref{eq:totalCE2}) is giving a penalty if the generated image on the whole domain does not look similar to a real turbulent configuration, by applying  the judgement of the (previously trained) discriminator D:
\beq
\Lcal_{adv} = \log(1-D[G[{\bf z}]])
\label{eq:lossCE2}
\eeq
where $\lambda_{adv} = 0.1$ is a parameter used to balance between the two losses and where the input data are updated to fool D, $-\partial_{{\bf z}} \Lcal_{adv}$. 
In Fig.~\ref{fig:CE2spotsvalidation} we show the equivalent of Fig.~\ref{fig:CE1spotsvalidation} for three different experiments using CE2. As one can see from data in column 5,  now there is a non zero  error, $\Delta_v(x_1)$ also outside the missing region because CE2 generate the flow in the whole domain. The overall performances have a maximal averaged error $\langle \Delta_v(x_1)\rangle$ of the order of $40\%$ roughly a factor 2 larger than CE1. It is important to stress  again that we didn't perform any systematic optimization of network structures and/or learning parameters being primarily interested here  to the proof of concept aspects. In Fig.~\ref{fig:physics2} we show the comparisons against the DNS (ground truth) for spectra, velocity and vorticity PDFs, again with  good overall results but  worse performance with respect to CE1, especially concerning the PDF of the local $L_2$ norm (right panel), which presents now a fatter right tail.  

\section{Equation-informed DA: Nudging}
\label{sec:Nudging}
In this section we confront the results obtained with CE1 and CE2 against what one can achieve using  Nudging \cite{Hoke76, Lakshmivarahan13, clark_di_leoni_synchronization_2020}, a  DA scheme popular among numerical weather forecast community and in economic fields \cite{benartzi2017should}. Nudging is distinguished from the methods presented above in two senses: it relies entirely on solving the equations of motion and it is also applicable to reconstruct trajectories in phase space. The scheme works by introducing a relaxation term to the equations of motion,  penalizing the flow when it deviates from a given reference state, set by the data one wants to reconstruct/assimilate. In rotating turbulence it takes the following form:
\begin{equation}
    \frac{ \partial \bm{v}}{\partial t} + \bm{v} \cdot \bm{\nabla} \bm{v} + 2 \Omega \hat{x}_3 \times \bm{v} = -  \bm{\nabla} p + \nu \nabla^2 \bm{v} + \gamma (1-\hat{M})_{x_3} \odot (\bm{v} - \bm{v}_{\rm ref})
    \label{nudging}
\end{equation}
where $\gamma=100$ is the intensity of the nudging, $(1-\hat{M})_{x_3} \odot$ is the same  filtering operator used in the previous section, replicated identically along the $x_3$ direction, such that  the nudging term acts only on the locations where the reference damaged flow, $\bm{v}_{\rm ref}$, is known.  The idea behind the method is simple, instead of using a machine to learn  (local and non-local) information in the 2d images we ask the NSE to provide the {\it semantic} background to fill the gaps, providing the whole 3d field in the whole volume $\bv(\bx,t)$ and for all times. Notice that NSE need to be evolved on the original volume size, even if the input is masked to be a subset of a 2d slice.  Notice also that usually Nudging is used to reproduce also temporal evolution, while here we will implement it to reconstruct one given, frozen in time, 2d velocity snapshot (which is replicated identically along the $x_3$).  Another important difference wrt Context Encoders is that we do not need any training section using hundred thousands examples from the ground truth as in ML applications, provided that the equations used are the ones that have also generated the corrupted images. It is also key to realize that we do not need to supply any external mechanical forcing mechanism to \eqref{nudging}, i.e. the reconstruction is obtained without knowing the external stirring. Nudging was shown to be able to reconstruct, with different degrees of accuracy, high Reynolds number three dimensional turbulent flows \cite{clark_di_leoni_synchronization_2020}, two dimensional flow \cite{Gesho16, Farhat16} and Rayleigh-Bernard convection \cite{Farhat19}. Moreover, Nudging can make up for inferring missing terms in the equation, such as forcing or even rotation \cite{Clark18}. In the following we show the performance of nudging to reconstruct missing hole in the middle of a 2d slice exactly as we did before for the two Context Encoders. We performed a long numerical integration of (\ref{nudging}) on the original box size, with  $256^3$ collocation points and supplying the 2d damaged input given by  $(1-\hat{M})_{x_3} \odot \cI_v$, where  $\cI_v$,  denoted here $\bv_{ref}$  to follow the notation used by the community,  is the three velocity components in a $64^2$ 2d  plane $(x_1,x_2)$. As said, the image is extended along the vertical direction in order to form a 3d volume (but no vertical fluctuations are introduced as we just stack up the same image). The resulting reference field for the nudging algorithm is constant in time. In this way we are not providing any extra information (vertical or temporal fluctuations) compared with the CE1 and CE2  cases discussed above. With this in mind, the result coming from the nudging is obtained as an average both  in time and along the vertical direction.

\begin{figure}
    \centering
    \includegraphics[width=1.0\textwidth]{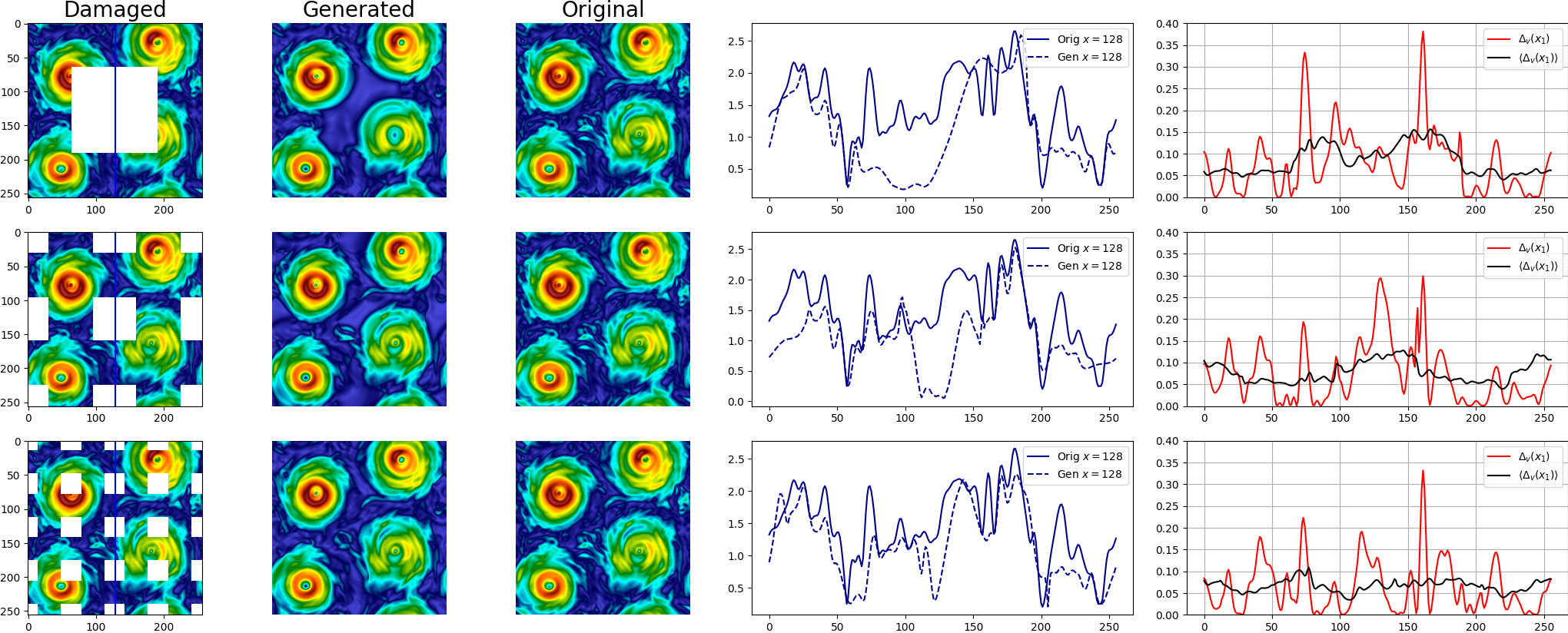}
    \caption{The same as in Fig.(\ref{fig:CE1spotsvalidation}) but for nudging. Here in the 2nd column we show the average in time and along the $x_3$ direction of the nudged-NSE (\ref{nudging}) output, $\langle |\bv (x_1,x_2)|^2\rangle_{t,x_3}$. In the 5th column, the black curve is obtained with a further average over three different nudging experiment.  }
    \label{fig:nudging}
\end{figure}

In Fig.~\ref{fig:nudging} we show the results of one of our experiments at changing $\ell/L$  following what done for the CE1 and shown in  Fig.~\ref{fig:CE1spotsvalidation}. As for CE2 case, nudging is introducing errors also where data are supplied, because NSE evolves in the whole (3+1)d domain. From the data shown in column 5, we can see that the averaged error $\langle \Delta_v(x_1) \rangle$ is  more noisy wrt to the CE1 and CE2 corresponding cases. This is due to the fact that we performed only three experiments with different nudging fields because of the much heavier computational overhead introduced by the need to solve NSE for each test. Nevertheless, we  notice that the overall errors are of similar order wrt to Figs.~\ref{fig:CE1spotsvalidation}-\ref{fig:CE2spotsvalidation}, although some small-scale reconstruction deficiencies can be spotted for the bigger gap  case  $\ell/L=1/2$ (first row).  
Some of  these problems could be mitigated if the reference field is not constant in time \cite{clark_di_leoni_synchronization_2020}. 
\section{Conclusions}
\label{sec:Conclusion}
We have presented a series of investigations meant to test the applicability of tools widely used by the CV community for image inpainting to turbulent data reconstruction  from the TURB-Rot database \cite{turbrot}. At difference from the typical set of images database where CV algorithms are trained, our case is made of a series of complex {\it scenes}, with random properties and multi-scale fluctuations.  We have {presented results for} two different Context Encoders, called  CE1 and CE2. CE1 is able to provide (once trained) a guess for the missing data without any further inference. Contrary, CE2 is based on a two step training, first the machine learns how to generate ungappy images and second it specialises to further fill the gap, needing a new training phase  for each new filling. Results among the two are  very satisfactory with CE1 performing better, even though a few caveats must be made. {First,  we didn't try to push the hyper-parameters settings to a level that cannot be improved further. Second, we do not think there exists ‘the best’ method to reconstruct and assimilate turbulent data. Turbulence has many common features but also many small subtle differences in different realization in nature and in the labs. So, the question ‘what is the best method to assimilate turbulent data’ is not well posed. Even for the same set-up, different methods can have different level of optimality depending on the quality and quantity of missing data, a trivial example being given by the case when data are missing in regions of size smaller than the Kolmogorov scale, where any nontrivial interpolation scheme would reconstruct pretty well. Similarly, all ML tools need to be retrained in the presence of big changes in the context: robustness and generalization being very limited. The scope of this paper is to present two different ML tools, and show that we have new tools in the box. Optimality and uniqueness of reconstruction cannot be proven rigorously due to the high dimensionality of the phase-space and the presence of  rough (multifractal) flow configuration.}
To this end we provide the whole database online and open such as to trigger the interested community to improve our results \cite{turbrot}. We found that the two CE algorithms are good also for {\it semantic} filling, i.e. when a big part of the image is missing, {with maximum average L2 error of the order of $10\%$ that must be considered good or bad depending on the other methods that one can apply. In our case, we do not know of any other attempt to try to reconstruct such a kind of complex flow as we did here and we consider the result highly non trivial.}  We found that large-scale features (velocity components) are reconstructed better than small-scale ones (vorticity) and that for the latter  it is key to provide the whole set of velocity components in input to improve the inpainting. {Point-to-point reconstruction in turbulence is theoretically limited by the presence of spatio-temporal chaos, which makes the configuration in any region (or time window) strongly sensitive to the boundary (or initial) conditions. As a result, it is not surprising that the vorticity field is less accurately reconstructed, given the very short correlation length of turbulent gradients. Statistical reconstruction is easier, and indeed our two CEs are performing well for this aspects, as shown by the middle panels of Figs. \ref{fig:physics}, \ref{fig:physicsomega}, \ref{fig:physics2} and in Figs. \ref{fig:physics_sgs} and \ref{fig:maxes}. It would be interesting to apply the same techniques to flows with a mean profile, to see if the presence of non-statistical information is helpful to condition also the fluctuating fields. } 

Finally, we present results using  Nudging, an equation-informed tool well established among the applied communities in numerical weather forecasts, geophysics and others. For our application, Nudging requires a new simulation of a 3d fully developed turbulent flows for each data reconstruction experiment and thus it is computationally much heavier than the two ML algorithms (if already trained).  {The trade-off is that Nudging will provide for a field reconstruction that is fully respectful of all symmetries and kinematic constraints enjoyed by the PDE, something that is not always easy to achieve with Convolutional Neural Networks. For instance, CE2 was trained without imposing periodic boundary conditions on the whole domain, a feature that the GAN needed to discover by itself. It will be interesting to study the trade-off  between simplicity of the implemented GAN architecture and the accuracy obtained by imposing known physical constraints \cite{raissi2019physics,wu2020enforcing,wang2019towards,erichson2019physics,lusch2018deep,kashinath2020enforcing,mohan2020embedding}}

Another advantage of Nudging is that it does not need training, it works even with  one damaged configuration only, it relies on the temporal evolution of the NSE to explore the phase space  and provide a realistic prior for the missing information (see \cite{pawar2020long} for a recent attempt to fuse nudging with recurrent neural network). It would be interesting to compare it with other ML tools that are also based on one single image analysis to infer the statistics of the missing region \cite{ulyanov2018deep,de2019data}. It is important to notice that the application we have selected, fully developed turbulence with rotation, presents highly non-trivial aspects, e.g. multi-scale non-differentiable fields with fluctuations over 2-3 decades, without a  mean flow, or boundaries. As a result, many techniques based on texturing \cite{efros1999texture}, non-linear interpolation \cite{shen2002mathematical} or modal analysis \cite{scherlrobust}, as well as patch methods that search similar patterns in the available image \cite{barnes2009patchmatch} could be in troubles to get the same accuracy. Moreover, we reconstruct only one slice of the whole 3d domain, so without any hint on the 3d features (or on temporal evolution). In conclusions, we have provided a first attempt for flow reconstruction  by using state-of-the-art context encoders based on Deep-GAN, and applying it to a paradigmatic {\it hard} problem for fluid dynamics. {Many  qualitative and quantitative   questions remain open, e.g. 
what happens when the damaged area  is irregular and multiscale? what about 3d reconstruction or (3+1)d when also time is entering in the measurements, or applications to PIV techniques~\cite{nobach2009limitations}? For the former, there is a huge literature in the CV community regarding inpainting under different quality of damages and the tools are pretty robust and flexible.  Similarly, there exists inpainting ML techniques for 3d images and videos, including temporal information also. E.g.,  to extend our idea to deal with 3D input configurations instead of 2D images, the initial part of the encoder in CE1 or the discriminator of CE2 network should be modeled using 3D CNNs, which are now mature enough to be computationally efficient and usable (see \cite{ji20123d,kpkl2019resource}). Exploiting temporal correlation is also a promising direction, typically hindered by the problem of large scale sweeping effects, calling for the need to follow fluid structures and extreme events in a quasi-Lagrangian reference frame.}
\\

When applying ML tools to turbulence, we need to use  quantitative benchmarks to assess the performance of different networks, being visual comparisons or the spectral properties obviously not enough for such complex physics. In particular, for DA and  tools there is the need to assess with local $L_2$ norm the goodness of the results, being interested to reconstruct  non-Gaussian fluctuations in the {\it correct position} and with the {\it correct intensity}. \\
We hope this study will trigger other systematic validation and application of ML algorithms to complex turbulent features. \\

\section*{ACKNOWLEDGMENTS}
This project has received partial funding from the European Research Council (ERC) under the European Union’s Horizon 2020 research and innovation programme (Grant Agreement No.882340).

\appendix
\section{CE1} 
\label{appdx:ACE1}
\begin{figure}
    \centering
    \includegraphics[width=1.0\textwidth]{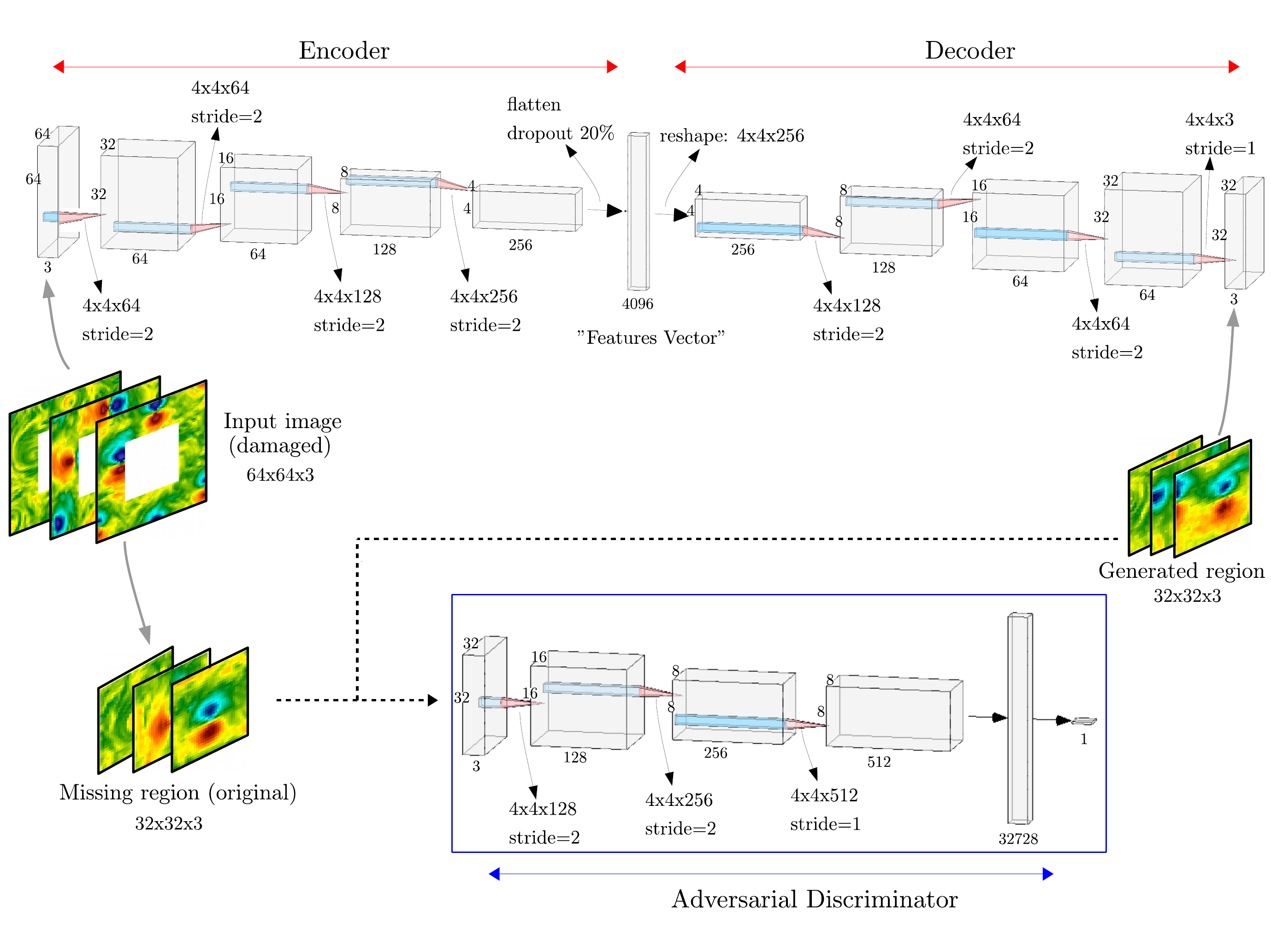}
    \caption{Sketch of the context encoder, CE1, architecture used for semantic inpainting with joint reconstruction and adversarial loss.  Here and in the following figure visualization is obtained using NN-SVG software, \cite{lenail2019nn}.}
    \label{fig:ACE1}
\end{figure}

\begin{figure}
    \centering
    \includegraphics[width=1.0\textwidth]{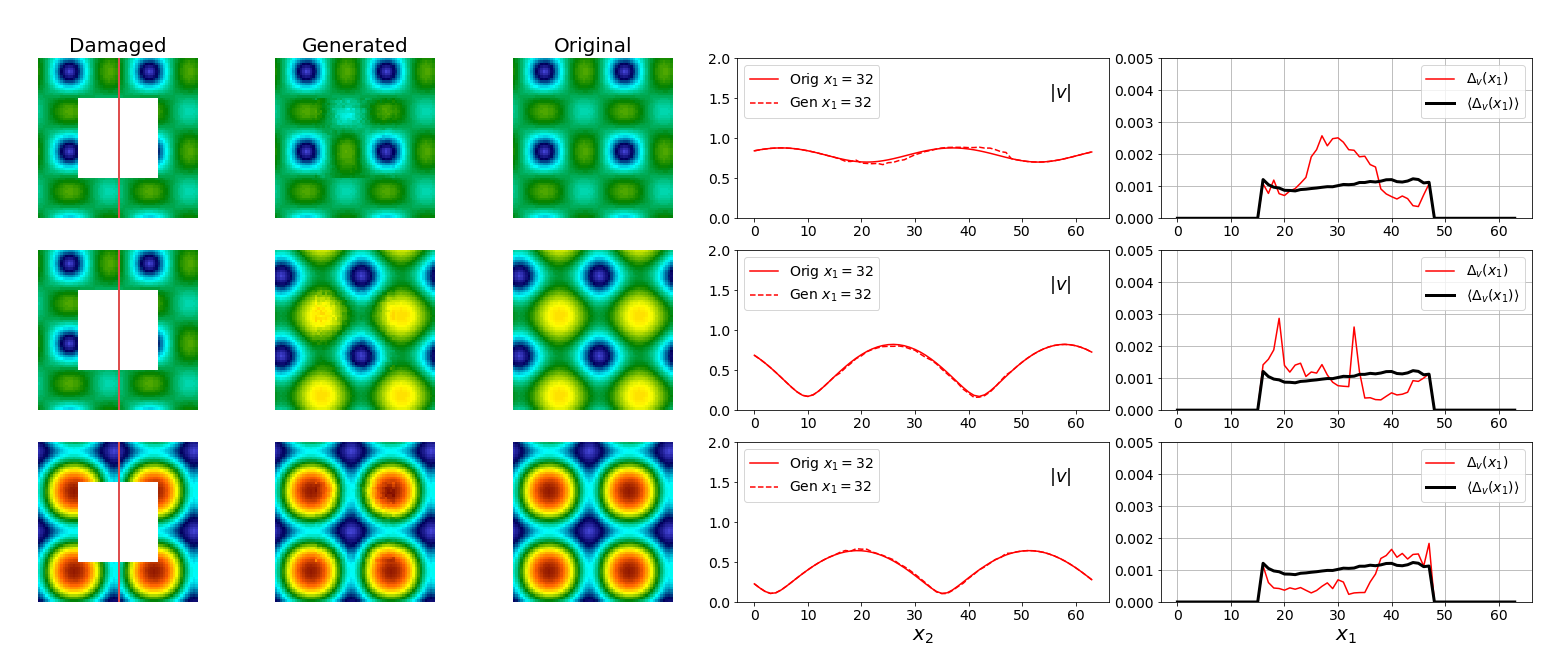}
    \caption{\PC{Benchmark of reconstruction on a 2d slice from a 3d random Taylor-Green flow. Interpolation is obtained with 1 part of one thousand precision in the whole region. } }
    \label{fig:TG}
\end{figure}

For the implementation of the CE1 we used TensorFlow \cite{tensorflow2015-whitepaper} libraries with a Python interface, optimized to run on GPUs. 
As introduced in Sec.~\ref{sec:CE1}, the overall architecture is composed by a Generator plus an adversarial Discriminator network.
The  Generator can be further divided into an encoder and a decoder part, see Fig.\ref{fig:ACE1}. The encoder is asked to identify all the relevant features of the three damaged images corresponding to the three velocity components. We have an input of size $64\times64\times3$, where we masked with $0$ values the gap region of size $\ell \times \ell$.
The encoder is based on several convolutional layers with stride $2$ that reduce the dimension of the input up to a $4\times4$ image. The last layer of the encoder, composed of $256$ convolutional filters,  is reshaped in a way to form a one-dimensional {\it features vector} of size $4096$ given in input to the decoder part of the generator. A $20\%$ dropout is added to the features vector such as to reduce over-fitting. The decoder architecture that completes the Generator part, is symmetric with respect to the encoder, and it is made of several layers of transposed convolutions such as to bring the features vector to the size of the original missing region, in our case of $32\times32\times3$.
A first evaluation of the quality of the generated patch is measured by the $L_2$ norm of the difference between the generated and the original portion of the image, as we shown in eq.~\eqref{eq:Lrec}. The quality of the generated image is generally blurred and to reach a sharper reconstruction a joint loss with an adversarial architecture is required \cite{pathak2016context}. The discriminator, shown at the bottom of Fig.\ref{fig:ACE1}, is asked to recognize the original from the generated image without looking at the context but only comparing the data in the missing region and the generated part \cite{pathak2016context}.
The discriminator architecture similarly to the encoder is based on several convolutional layers that transform the input of size $32\times32\times3$ to a one-dimensional vector that is then transformed by a fully connected layer and a softmax activation function to the binary probability of the input image to be original or generated. All convolutional layers in the network are connected with LeakyRELU activation functions with a coefficient of $0.2$. The total loss defined in eq.~\eqref{eq:original} is then constructed by weighting the reconstruction and the adversarial terms with the following factors, $\lambda_{rec} = 0.999$ and $\lambda_{adv} = 0.001$.
From the evolution of the two losses during a typical training experiment as shown in Fig.~\ref{fig:CE1training}, we can see that $\mathcal{L}_{adv}$ is almost two orders of magnitude higher than $\mathcal{L}_{rec}$, hence from the choice of $\lambda_{adv}$ and $\lambda_{rec}$ discussed above we can conclude that the discriminator loss was found to be optimal when its weight is around $10\%$ of the total loss $\mathcal{L}$.
It is important to stress that to achieve a good training with the factors discussed above, we have normalized the input data to be in the $[-1;1]$ range, namely we rescaled the intensity of each pixel  by subtracting first the mean, $m = \frac{\max(\mathcal{I}_v)+\min(\mathcal{I}_v)}{2}$, and then normalising by  $\delta = \frac{\max(\mathcal{I}_v)-\min(\mathcal{I}_v)}{2}$, where the maximum and the minimum are calculated over the whole training set of size $N_{tra}$. The Generator output was then shifted and rescaled back to match the original range of the velocity fields. 
To reach a better balance  between the Generator and the Discriminator parts we used a learning rate for G $2$ times higher than that of the D. In particular the context encoder was trained with a learning rate of $2\times10^{-4}$ while the discriminator learning rate was fixed at $1\times10^{-4}$.
The optimizer used to train the network is ADAM with parameters  $\beta_1=0.5$, $ \beta_2=0.999$ and $\epsilon=10^{-8}$, see \cite{kingma2014adam}.
The mini-batch used in our training is composed of $128$ images. Hence, because $N_{tra}=81920$, each epoch consists of $640$ mini-batches. After every mini-batch is analysed the network weights are updated back-propagating the mean error estimated on the mini-batch. {For the application to case (ii) of Sec.~\ref{sec:fr2} the structure of the network is the same except that in input and output we have one image  instead of three.}
\PC{A first validation for the reconstruction of 2d slices of a simple random 3d Taylor-Green flow is shown in Fig. \ref{fig:TG}. Being the flow differentiable and not-multiscale, the reconstruction task is achievable with a non-linear interpolation. In the figure we show that the network  CE1 is able to fulfill the requirement within 1 part over 1 thousand even for the largest gap. }
\section{CE2}
\label{appdx:ACE2}
\begin{figure}
    \centering
    \includegraphics[width=1.0\textwidth]{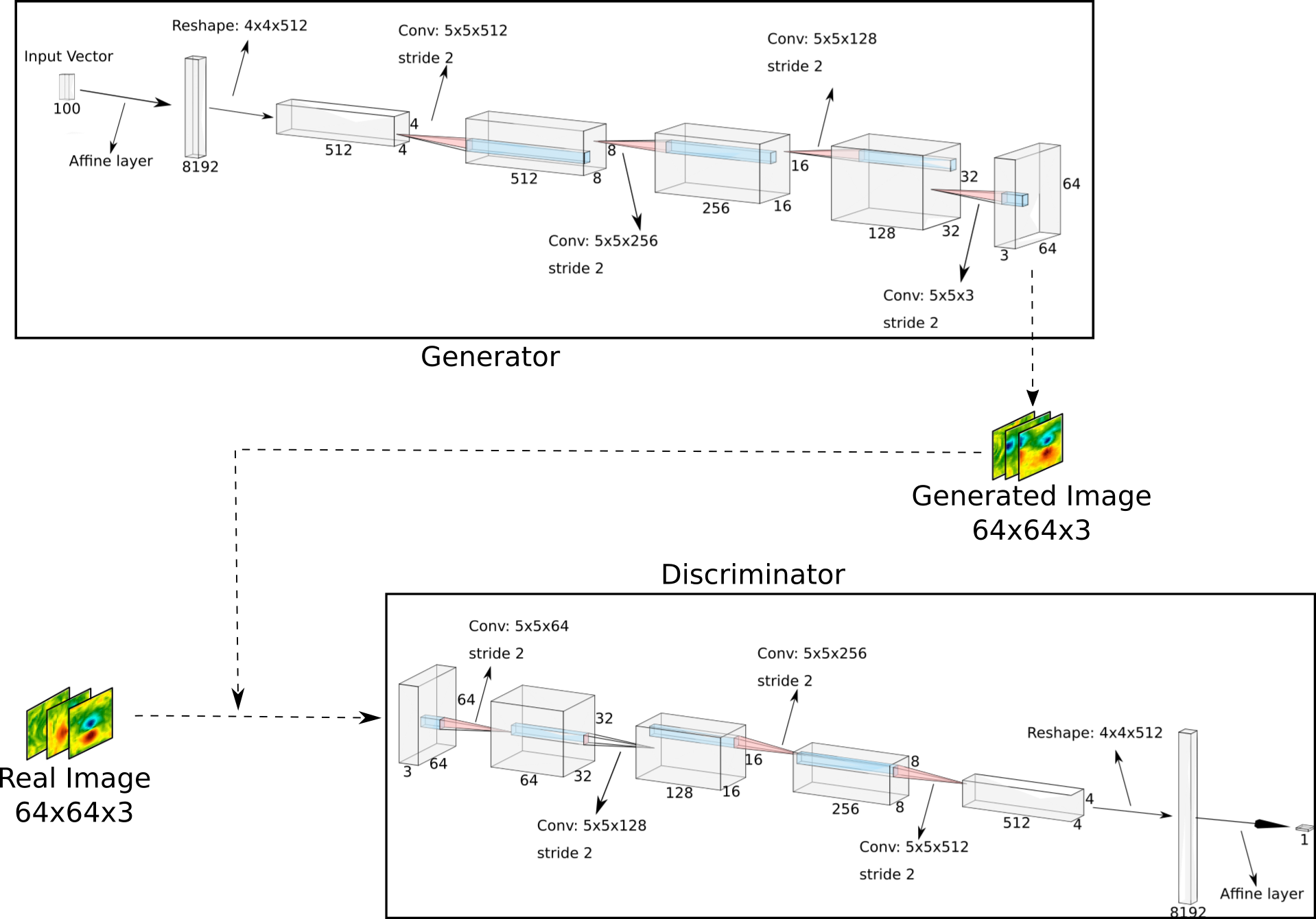}
    \caption{Structure of the CE2 network: the generator from a 100-dimensional input vector produces a 64x64x3 image, the discriminator decides if an input 64x64x3 image is coming from real data or is generated. }
    \label{fig:ACE2}
\end{figure}

As introduced in Sec.~\ref{sec:CE2}, the CE2 architecture is a GAN composed by a Generator plus an adversarial Discriminator network, as shown in Fig.\ref{fig:ACE2}.
{The networks largely derive their structure from the VGG (and AlexNet) CNNs, which proved very successful in the ImageNet image classification. The initial part of VGG is made of a sequence of convolutions of filter size=3 and is the key ingredient to infer underlying structural features from 2D images. We tried variations of this schema changing the filter size to 3,4 or 5, retaining the best configuration, that is the one we propose.}
We started from the code in \cite{amos2016image}, which is a python program using the TensorFlow library, with minor changes to adapt it to our  dataset.
In detail, the generator part is a network aimed to produce a 3-dimensional velocity field starting from an arbitrary vector, { ${\bf z}$, with $T$-real entries ($T=100$ in our implementation) } which should prescribe the desired features. {The feature size of 100 was chosen as a compromise between accuracy versus training time after attempts with $T=300, 800, 1000$ that did not improve the error in a significative way. This is in line with current impainting algorithms for images of similar input size.}
It is similar to the decoder part of the CE1, {plus the insertion} of  a fully connected layer from the input to an intermediate layer of size $8192$ that then is reshaped as $4\times4\times512$ volume. The latter is used as the first stage of the next layers: with four successive transposed convolutions of size $5 \times 5$ with stride=$2$ (with ReLu activation) we go from $4\times4\times512$ to $8\times8\times512$ to $16\times16\times256$ to $32\times32\times128$ to finally obtain, using $3$  transposed convolutions, the output velocity field of size $64\times64\times3$.
The Discriminator network, takes a 3 component velocity field of dimension $64 \times 64$ and outputs the probability that the input is real. Similar to the analogous in the CE1 network, using convolutional layers of size $5 \times 5$ with stride=$2$ (with LeakyRELU activation) we reduce the dimension of the input down to a $4 \times 4$ image with 512 channels. After reshaping to a vector of size $8192$, the last stage is made of a fully connected layer and a sigmoid activation function, providing the probability to be real or created by the generator.\\
The generator component of the network is trained to {\it minimize} the function: 
\begin{equation}
  \Lcal_{gen} =  \mathbb{E}_{{\bf z}} \{ \log (1 - D[G[{\bf z}]]) \}
  \label{eq:ce2_genCost}
\end{equation}
in order to fool the discriminator with its fake output.
The discriminator component of the network is trained to {\it maximize} the function: 
\begin{equation}
  \Lcal_{adv} =  \mathbb{E}_{\cI_v, {\bf z}} \{ \log(D[\cI_v]) + \log(1-D[G[{\bf z}]]) \}
  \label{eq:ce2_discrCost}
\end{equation}
in order to assign probability $1$ to real fields and probability $0$ to generated fields.
As with CE1, we have normalized independently the $3$ components of the input data to be in the $[-1:1]$ range. The Generator output was then re-scaled to match the original range of the velocity fields. Both networks are trained with the ADAM optimizer, \cite{kingma2014adam}, using a learning rate of $2\times10^{-4}$, $\beta_1=0.5$, $ \beta_2=0.999$ and $\epsilon=10^{-8}$. 
The training was performed using a batch size equal to 256, so the $81 \times 1024$ planes of the dataset were divided into 324 minibatches for a total of 100 Epochs.
After the Deep-Gan is trained, its weights are frozen and used for the reconstruction phase: to recover the optimal ${\bf z}^*$-encoding that is closest to each corrupted image, the loss given in eq. \ref{eq:totalCE2} is minimized by back propagation to the input data using an ADAM optimization. In our test, we used a set of N=1024 images to be completed (selected outside the train dataset for the GAN), each minimized independently (batch size=1), with  a learning rate of $2\times10^{-2}$, $\beta_1=0.9$, $ \beta_2=0.999$ and $\epsilon=10^{-8}$ for a high number of epochs (10 000).

\section{Database}
\label{appdx:DataBase}
\begin{figure}
    \centering
    \includegraphics[width=0.8\textwidth]{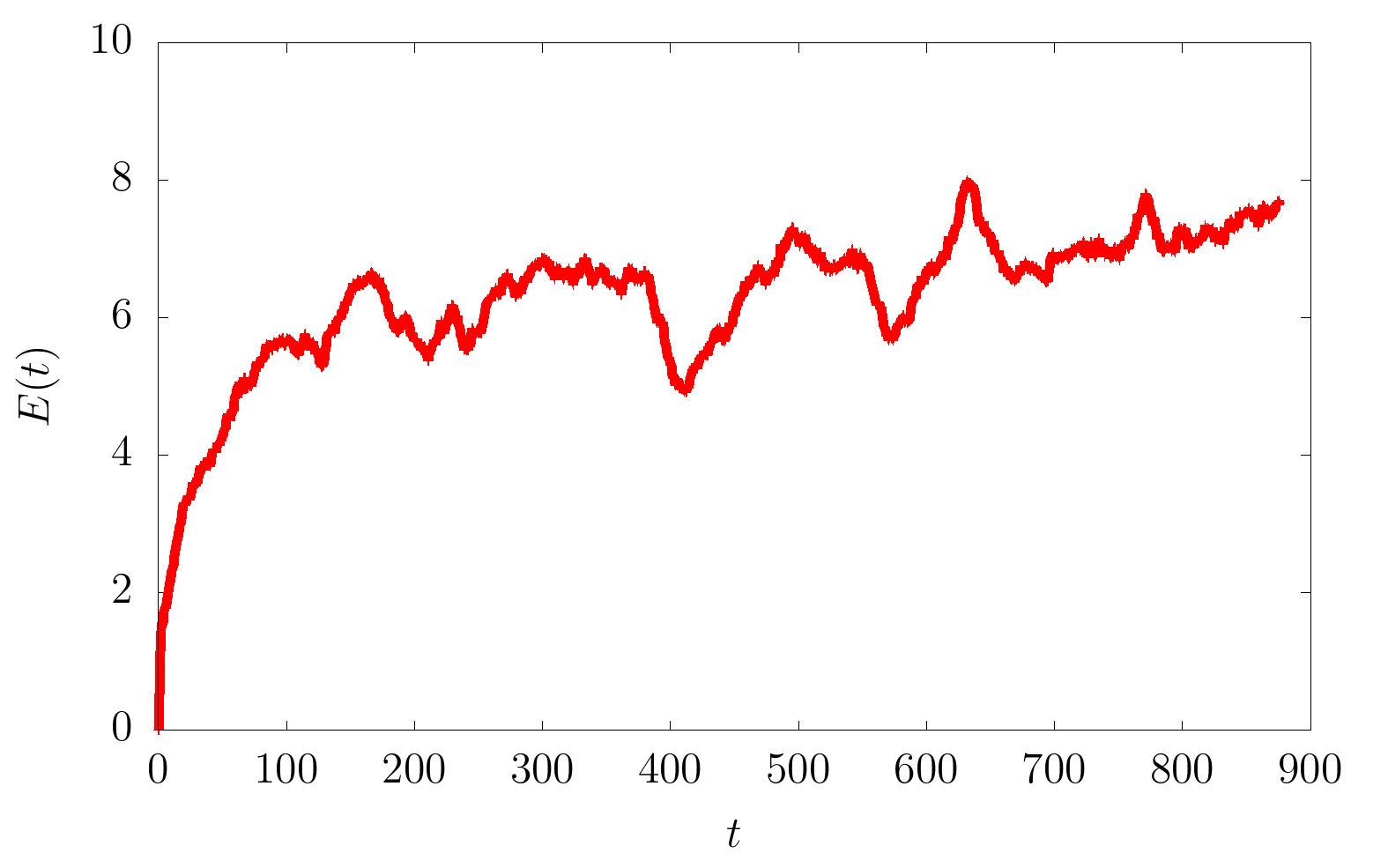}
    \caption{Energy evolution for the turbulent flow generated during the simulation performed to generate the database of $600$ different configuration at a resolution of $256^3$ grid points. The velocity fields are extracted from time $t=276$, to reach the flow thermalization, up to the final time of $t=876$ every $t = 1$ which corresponds to $2000$ simulation time-steps.}
    \label{fig:ene_evol}
\end{figure}

Data are key for machine learning; hence, our aim to produce and curate benchmark datasets (and software) to foster interest among researchers in related fields. Complex flows and complex fluids  benchmarks are more challenging than the {\it traditional} image datasets found in computer vision and machine learning: our data have high-resolution, multi-scale features, are inherently stochastic, without long term spatial and temporal correlations, and -often- with multi-scale non differentiable properties.\\
We share the believe that we are entering a new era in fluid mechanics research. Decades (centuries)  of great theoretical, numerical and experimental developments based on first principles, brute-force simulations of the equations of motion or accurate observation of nature are now confronting  with data-driven tools and analysis.\\
The database used is SMART-Rot a subset of data deployed in Smart-TURB \cite{turbrot}  and it is  prepared in four steps:
\begin{itemize}
    \item The DNS simulations at a resolution of $256^3$ grid points are truncated in Fourier space retaining till $k_f \le  32$, then transformed back in real space downscaling them to $64^3$
    \item From the whole time evolution, a number of $600$ snapshots is used, with large temporal separation to decrease correlations (see Fig.~\ref{fig:ene_evol}).
    \item For each  configuration, 16 horizontal cuts in the plane $(x_1,x_2)$ are selected. To increase the dataset, we shift each plane in $11$ different random ways such as to preserve periodic boundary conditions and to obtain a total of $16 \times 11 = 176$ planes
    \item Finally, the dataset is composed of $600 \times 176 = 105600$ planes, which are organized in a random order to avoid any time-correlation between successive planes.
\end{itemize}  
The database of the configurations was then divided into train set, the first $80 \times 1024$ planes, and validation set, the next $20 \times 1024$ planes.\\
The full database TURB-Rot is made public, available for download using the SMART-Turb portal \cite{turbrot}.
The portal uses the concept of "Dataset" to aggregate resources related to the same simulation: we have released both the original full resolution of 3d DNS snapshot at $256^3$ grid points and the database of $105600$ ($x_1,x_2$)-planes with size $64 \times 64$ used by CE1 and CE2, in a dataset named $\mathbf{TURB-Rot}$. 

\newpage

\bibliographystyle{unsrt}
\bibliography{biblio}

\end{document}